\documentclass[preprint,12pt]{aastex}
\usepackage{epsfig}
\usepackage{natbib}
\usepackage{amssymb}
\usepackage{amsbsy}
\usepackage{natbib}
\usepackage{subfigure}
\usepackage[mathcal]{euscript}
\usepackage{float}
\usepackage[showframe=false]{geometry}
\usepackage{changepage}
\usepackage{rotating}
\usepackage{xcolor}
\bibpunct{(}{)}{;}{a}{}{,}

\begin{document}
\title{Rapid Rotation of Low-Mass Red Giants Using APOKASC: A Measure of Interaction Rates on the Post-main-sequence}
\author{
Jamie Tayar\altaffilmark{1}, 
Tugdual Ceillier\altaffilmark{2}, 
D.~A.~Garc{\'i}a-Hern{\'a}ndez\altaffilmark{3,4}, 
Nicholas W.~Troup\altaffilmark{5}, 
Savita Mathur\altaffilmark{6}, 
Rafael A.~Garc{\'i}a\altaffilmark{2}, 
O.~Zamora\altaffilmark{3,4}, 
Jennifer A.~Johnson\altaffilmark{1,7}, 
Marc H.~Pinsonneault\altaffilmark{1}, 
Szabolcs~M{\'e}sz{\'a}ros\altaffilmark{8}, 
Carlos Allende Prieto\altaffilmark{3,4}, 
William J.~Chaplin\altaffilmark{9,10}, 
Yvonne Elsworth\altaffilmark{9,10}, 
David L.~Nidever\altaffilmark{11}, 
David Salabert\altaffilmark{2}, 
Donald P.~Schneider\altaffilmark{12,13}, 
Aldo Serenelli\altaffilmark{14}, 
Matthew Shetrone\altaffilmark{15}, and 
Dennis Stello\altaffilmark{10,16} 
}
\altaffiltext{1}{Department of Astronomy, Ohio State University, 140 W 18th Ave, OH 43210, USA}
\altaffiltext{2}{Laboratoire AIM, CEA/DSM-CNRS, Universit\'e Paris 7 Diderot, IRFU/SAp, Centre de Saclay, 91191, Gif-sur-Yvette, France}
\altaffiltext{3}{Instituto de Astrof\'{\i}sica de Canarias (IAC), V\'{\i}a Lactea s/n, E-38205 La Laguna, Tenerife, Spain}
\altaffiltext{4}{Departamento de Astrof\'{\i}sica, Universidad de La Laguna (ULL), E-38206 La Laguna, Tenerife, Spain}
\altaffiltext{5}{Department of Astronomy, University of Virginia, Charlottesville, VA 22904-4325, USA}
\altaffiltext{6}{Space Science Institute, 4750 Walnut Street Suite 205, Boulder, CO 80301, USA}
\altaffiltext{7}{Center for Cosmology and Astro-Particle Physics, Ohio State University, Columbus, OH, United States}
\altaffiltext{8}{ELTE Gothard Astrophysical Observatory, H-9704 Szombathely, Szent Imre herceg st. 112, Hungary}
\altaffiltext{9}{School of Physics and Astronomy, University of Birmingham, Birmingham B15 2TT, UK}
\altaffiltext{10}{Stellar Astrophysics Centre, Department of Physics and Astronomy, Aarhus University, Ny Munkegade 120, DK-8000 Aarhus C, Denmark}
\altaffiltext{11}{Department of Astronomy, University of Michigan, Ann Arbor, MI, 48104, USA}
\altaffiltext{12}{Department of Astronomy and Astrophysics, The Pennsylvania State University, University Park, PA 16802}
\altaffiltext{13}{Institute for Gravitation and the Cosmos, The Pennsylvania State University, University Park, PA 16802}
\altaffiltext{14}{Institute of Space Sciences (CSIC-IEEC), Campus UAB, Bellaterra, 08193, Spain}
\altaffiltext{15}{Hobby-Eberly Telescope, 32 Fowlkes Rd, McDonald Observatory, Tx 79734-3005}
\altaffiltext{16}{Sydney Institute for Astronomy (SIfA), School of Physics, University of Sydney, NSW 2006, Australia}

\begin{abstract}
We investigate the occurrence rate of rapidly rotating ($v\sin i$$>$10 km~s$^{-1}$), low-mass giant stars in the APOGEE-\textit{Kepler} (APOKASC) fields with asteroseismic mass and surface gravity measurements. Such stars are likely merger products and their frequency places interesting constraints on stellar population models. We also identify anomalous rotators, i.e. stars with 5 km~s$^{-1}$$<$$v\sin i$$<$10 km~s$^{-1}$ that are rotating significantly faster than both angular momentum evolution predictions and the measured rates of similar stars. Our data set contains fewer rapid rotators than one would expect given measurements of the Galactic field star population, which likely indicates that asteroseismic detections are less common in rapidly rotating red giants. The number of low-mass moderate (5-10 km~s$^{-1}$) rotators in our sample gives a lower limit of 7\% for the rate at which low-mass stars interact on the upper red giant branch because single stars in this mass range are expected to rotate slowly. Finally, we classify the likely origin of the rapid or anomalous rotation where possible. KIC 10293335 is identified as a merger product and KIC 6501237 is a possible binary system of two oscillating red giants.

\end{abstract}
 
\section{Introduction}
Almost half of all low-mass stars form in multiple systems \citep{Raghavan2010}. Stars with companions in sufficiently close orbits can interact and produce important classes of objects such as blue stragglers, cataclysmic variables, low-mass white dwarfs, and Type Ia supernovae.  Binary interactions can induce strong internal mixing, alter nucleosynthetic yields, and even trigger stellar detonations.  {While some binary products are easily detectable, \citep[e.g. FK Comae stars,][]{BoppStencel1981} others are not, and} the rate at which binary interactions occur is highly sensitive to the poorly-known distribution of mass ratios and separations {\citep{DucheneKraus2013}}. Because stars are most likely to interact as they expand on the giant branch, placing a lower limit on the rate of stellar interactions during giant branch evolution offers an avenue to investigate this problem.

One way of identifying stars that have interacted or merged with a companion is measuring their rotation rates. A stellar interaction or merger can produce a star with a wide range of rotation rates at any mass and evolutionary state, because  orbital angular momentum can be exchanged with spin angular momentum during such an interaction { \citep[see, for further discussion of this process, ][]{Peterson1984, Mathys1991, LeonardLivio1995, Sills1997, Sills2001}}. Single stars, in contrast, have predictable rotation rates that depend mainly on mass, age, and evolutionary state. In some regimes, the main sequence angular momentum content of a single star is so small that any measurable rotation on the giant branch requires an interaction.

Given that some giant stars are observed to be rotating abnormally fast, three mechanisms have been suggested for creating rapid rotation in red giants: tidal interactions with a close companion, mergers, and accretion of material from a sub-stellar companion. Much work has been done to understand the frequency of binary systems and the rates at which such systems interact. \citet{Raghavan2010} suggest that 44\% of F through K stars form in multiple systems. Analysis of blue stragglers indicates that at minimum between 0.5\% and 4\% of binary systems interact on the main sequence to produce a remnant larger than the turnoff mass \citep{Sollima2008}. Analysis by \citet{Carlberg2011}, which combines the binary fraction of \citet{Famaey2005} and the period distribution derived by \citet{DuquennoyMayor1991}, indicates that between 1 and 2\% of K giants should be rapidly rotating ($v\sin i$ $>$ 10 km~s$^{-1}$) on the giant branch as a result of interaction with a companion.

Red giants could be similarly spun up if they consume a planetary companion \citep[see e.g.][]{Peterson1983, Soker2004, Massarotti2008, Carlberg2009}. Models indicate that a few Jupiter masses of material must be ingested to increase the rotation of a 1 M$_\sun$ star with a radius of 10 R$_\sun$ to more than 8 km~s$^{-1}$ \citep{Carlberg2009}. {As conservation of angular momentum dictates that the surface velocity must slow as the star expands, such systems would most likely appear as a concentration of} rapidly rotating stars on the lower first ascent red giant branch \citep{Carlberg2009}. Because the cross-section for interaction is largest at the tip of the giant branch, we also expect an enhancement of the number of rapid rotators in the red clump. Searches for stars that have been spun up by the accretion of planetary material have thus far been inconclusive \citep{Carlberg2011, Carlberg2013, Adamow2012}.

Consistent with model predictions of slow rotation, previous analyses have found that around 98\% of stars rotate slowly on the red giant branch \citep{Carlberg2011, deMediros1996}. These investigations therefore use rapid rotation on the giant branch to identify stars with an unusual history. Such studies define a single cutoff velocity between 8 and 12 km~s$^{-1}$ \citep{Drake2002,EiffReiners2011}, and declare all giants rotating faster than this rate to be 'rapid rotators'. While this is a conservative approach to selecting interaction products, a single value does not allow identification of all stars in a sample that have been spun up by unusual evolution, because low-mass stars would need a larger increase in velocity than intermediate-mass stars to cross any single velocity threshold. Additionally, using a single threshold value fails to identify stars rotating at moderate but measurable rates that in some mass regimes are likely to result from interactions. In order to identify a larger fraction of the stars spun up by interactions, we therefore take advantage of the well characterized APOKASC sample of stars, which involves both uniform spectroscopic analysis and measurements of seismic masses, to avoid having to choose a single threshold value for rapid rotation. We run stellar models to quantify the expected rotation rates of giant stars as a function of their physical properties and then use the available APOKASC seismic data to, for example, separate low-mass, low-metallicity stars from intermediate-mass stars with similar colors and to separate core helium burning clump stars from shell hydrogen burning stars of similar luminosity. This procedure allows us to both identify anomalously rotating objects that might normally have been missed and to classify, in some cases, the origins of the anomalous rotation.

In this work, we use the  APOGEE- \textit{Kepler} combined data set published in the APOKASC catalog \citep{Pinsonneault2014}. The spectroscopic properties in this catalog are produced by the Apache Point Observatory Galaxy Evolution Experiment, (APOGEE; Majewski et al. 2015 {in prep.}) a Sloan Digital Sky Survey III project \citep[SDSS-III;][]{Eisenstein2011} operating on the Sloan 2.5 meter telescope \citep{Gunn2006}. APOGEE has acquired over half a million high (R $\sim$ 22,500) resolution infrared spectra \citep{Wilson2012}. Its sample contains more than 1900 giant stars in the \textit{Kepler} field \citep{Borucki2010}, including seismically oscillating giants, eclipsing binaries, and planetary host stars \citep{Zasowski2013}. The APOGEE Stellar Parameters and Chemical Abundances Pipeline \citep[ASPCAP; Garcia Perez et al 2015, with cluster calibrations by][] {Meszaros2013} has performed an automated analysis of the spectra of these stars, comparing each observation to a library of non-rotating synthetic spectra to determine the star's composition, effective temperature, and metallicity.

The \textit{Kepler} satellite has produced photometric time series data of an unprecedented number of giants at millimagnitude precision. Asteroseismology, the use of the oscillation frequencies of a star to understand the underlying structure, has been applied to these data to determine surface gravities, masses, and radii for more than 13000 giants \citep{Stello2013}. The derived masses are further refined by comparison with stellar models \citep{Pinsonneault2014}.  We combine information from \textit{Kepler} and ASPCAP in our identification of rapidly rotating stars and our exploration of the probable causes for this rapid rotation.

This paper is organized as follows: in Section 2, we detail the data set and explain the calculation of rotational broadening. In Section 3 we model the expected giant branch rotation rates. In Section 4, we enumerate the rapidly rotating stars and discuss trends with mass, composition and evolutionary state.  We discuss suggested explanations for the observed anomalous rotation in Section 5 and conclude with constraints placed on upper red giant branch interaction rates in Section 6.

\section{Data Analysis}

Fundamental stellar parameters were taken from the APOKASC catalog \citep{Pinsonneault2014} based on {Sloan Digital Sky Survey Data Release 10 \citep{Ahn2014}.} To determine the projected surface rotational velocities of the {1950 APOKASC stars analyzed}, we used both the combined spectrum of each star {\citep{Holtzman2015}} and the best-fit template spectrum generated by the ASPCAP pipeline, {which is constructed assuming a Gaussian line spread function at a resolution of R=22,500}. {The combined observed spectra are corrected for telluric absorption and the vast majority of the spectra in our sample have signal-to-noise greater than 100.} The spectra are divided into three wavelength sections {corresponding to each of the three APOGEE detectors} and each section is individually compared to its own unbroadened template. As shown in Figure 1, by comparing the width of the cross-correlation peak between the observed spectrum and its template to the width of the cross-correlation peak between the template and an artificially rotationally broadened version of the template, a rotational broadening is determined \citep[see][with additions by Will Fischer and Chelsea Sharon]{ WhiteHillenbrand2004}. Spuriously high measurements of broadening (individual measurements more than five times the values measured for the other two wavelength sections) {arose in a few cases where the template spectrum was already broader than the observed spectrum or did not fit well. These values} were removed and the minimal possible measurement (2 km~s$^{-1}$) was substituted for that individual measurement. {As this occurred only in stars with no significant broadening, removing spurious values and averaging the two remaining measurements does not alter our published velocities.} Broadening due to microturbulence is included in the template spectra {as a linear function of the surface gravity}, and we verified that the effect of altering the limb darkening coefficient {from our adopted value of 0.60 to 0.25 {\citep[a value which has been suggested to be more appropriate for the H-band,][]{Howarth2011}}} is small ({ decreases by} $<$0.3 km~s$^{-1}$). We expect macroturbulent broadening in giants on the level of 5 to 10 km~s$^{-1}$ \citep{Lambert1987, Carney2008}, but because macroturbulence produces a more cuspy profile than rotational broadening \citep{Gray1992}, accounting for macroturbulent broadening only reduces our measured rotation rates by about 10\%, or about 1 km~s$^{-1}$ for the rapid rotators. We therefore assume that the measured line broadening corresponds to rotational broadening. Because there are three wavelength regions, we obtain three individual measurements of the rotational broadening; these are combined to determine a mean and standard deviation for the projected rotational velocity {which is presented in Table 1.}

As we did not determine the rotational broadening until after the spectral parameters of the stars had already been measured, it is possible that our measurements of rotation and other stellar parameters could be biased by rotation-dependent mismatches between the actual and unbroadened spectral features. Using the spectral libraries {\citep{Zamora2015}} developed for Data Release 12 analysis {\citep{Alam2015}} but including rotation broadening as a dimension in the ASPCAP fit decreases the measured velocity broadening by about 1 km~s$^{-1}$ on average and the stellar effective temperature by about 20 K (see Troup et al. 2015{, in prep.}). There also appears to be a velocity dependent offset in the measured metallicity which can be as large as 0.3 dex for a star rotating at 10 km~s$^{-1}$.  Given that systematic effects could alter our measured rotation rates at the 1 km~s$^{-1}$ level, we compare with surface rotation rates measured from higher resolution spectra of several APOGEE targets to tie our measurements to a fundamental rotation scale and remove such systematics.

\begin{figure}[H]
 \centering
\includegraphics[width=0.9\textwidth]{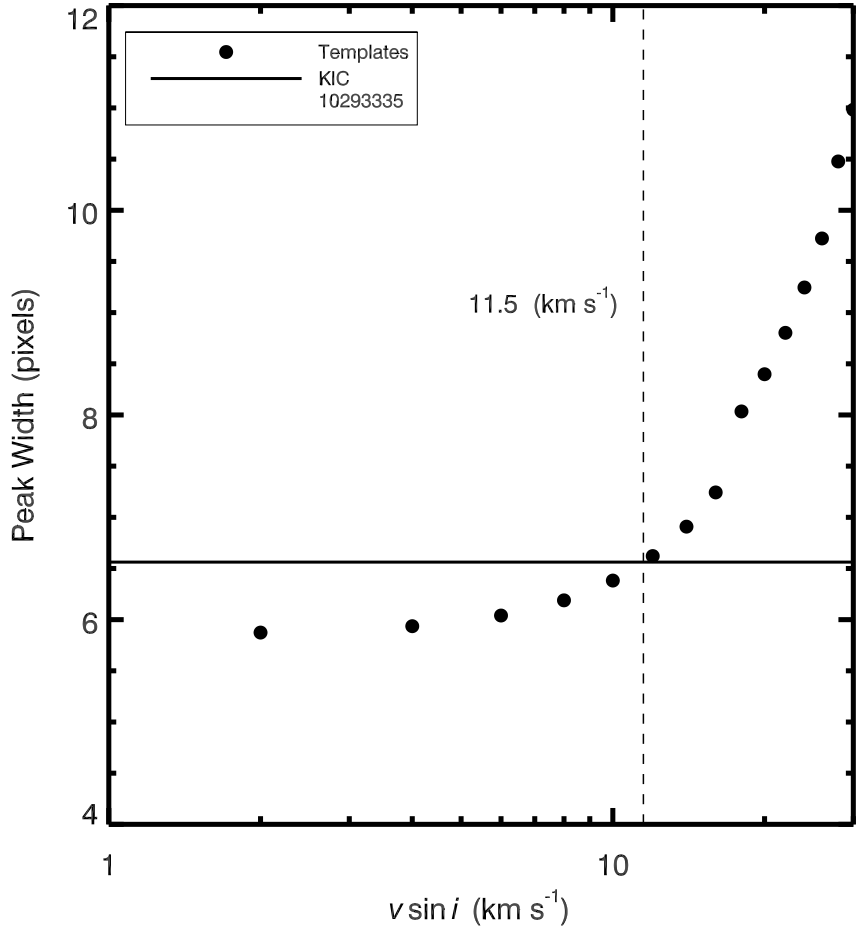}
\caption{The empirical full width at half maximum  of the cross correlation peak between the bluest section of the template spectrum of KIC 10293335 and a spun-up version of this template spectrum as a function of the rotational velocity (circles). The horizontal line represents the measured width of the cross correlation peak between the template spectrum and the observed spectrum, {and a linear interpolation between the empirical peak widths gives a $v\sin i$ of $\sim$ 11.5 km~s$^{-1}$).}}
\end{figure}

To control for template mismatches and other effects, as well as to calibrate our measured rotational broadenings, we identified an overlap of approximately 200 stars in the APOGEE sample with the analyses of high resolution (21,000 $<R<$ 80,000) optical spectra done by \citet{Molenda-Zakowicz2011, Molenda-Zakowicz2013}, \citet{Thygesen2012}, and \citet{Bruntt2012}. {In addition to our 1950 red giants, we also computed rotational velocities of these $\sim$200 dwarfs, subgiants, and giants.} Figure 2 compares the values derived using our method to $v\sin i$ values derived using these higher resolution spectra. While surveys working at higher spectral resolution are able to measure line broadenings below 5 km~s$^{-1}$, {reduced $\chi^2$ values indicate that} our results begin to differ from published results below this value; thus our measurements below 5 km~s$^{-1}$ should be considered nondetections. Above this value, our results are consistent with the Molenda-{\.Z}akowicz results, and are on average 4.1 km~s$^{-1}$ lower than the Bruntt $v\sin i$ values. We suggest that this discrepancy could be due either to a difference in velocity calibration or the fact that the Bruntt sample consists entirely of dwarfs and subgiants, regimes for which the APOGEE stellar parameters have not yet been calibrated.

\begin{figure}[H]
 \centering
\subfigure{\includegraphics[width=.32\textwidth, clip=true, trim=0.5in 1.5in 1.5in 1.6in] {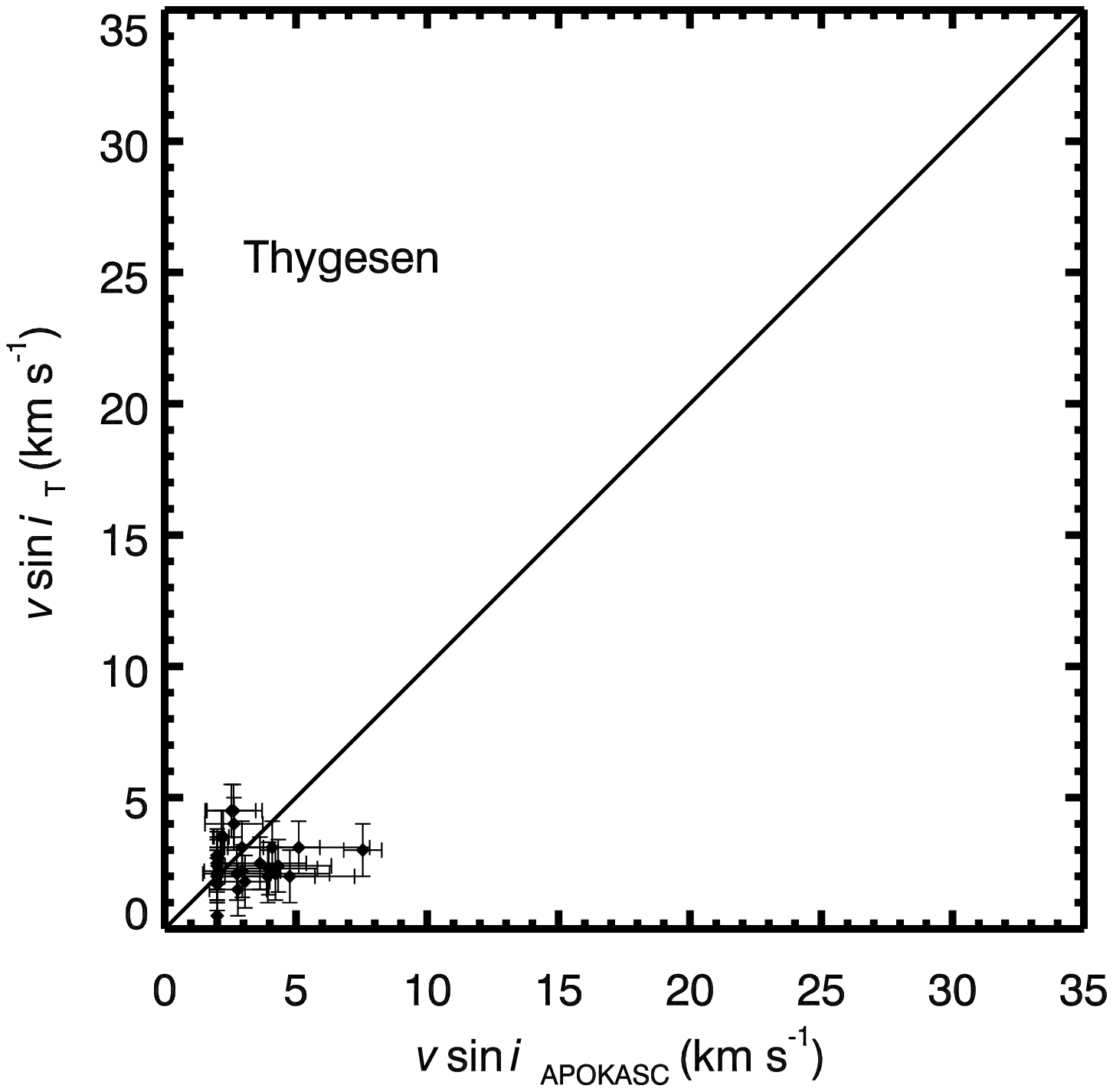}}
\subfigure{\includegraphics[width=.32\textwidth, clip=true, trim=0.5in 1.5in 1.5in 1.6in]{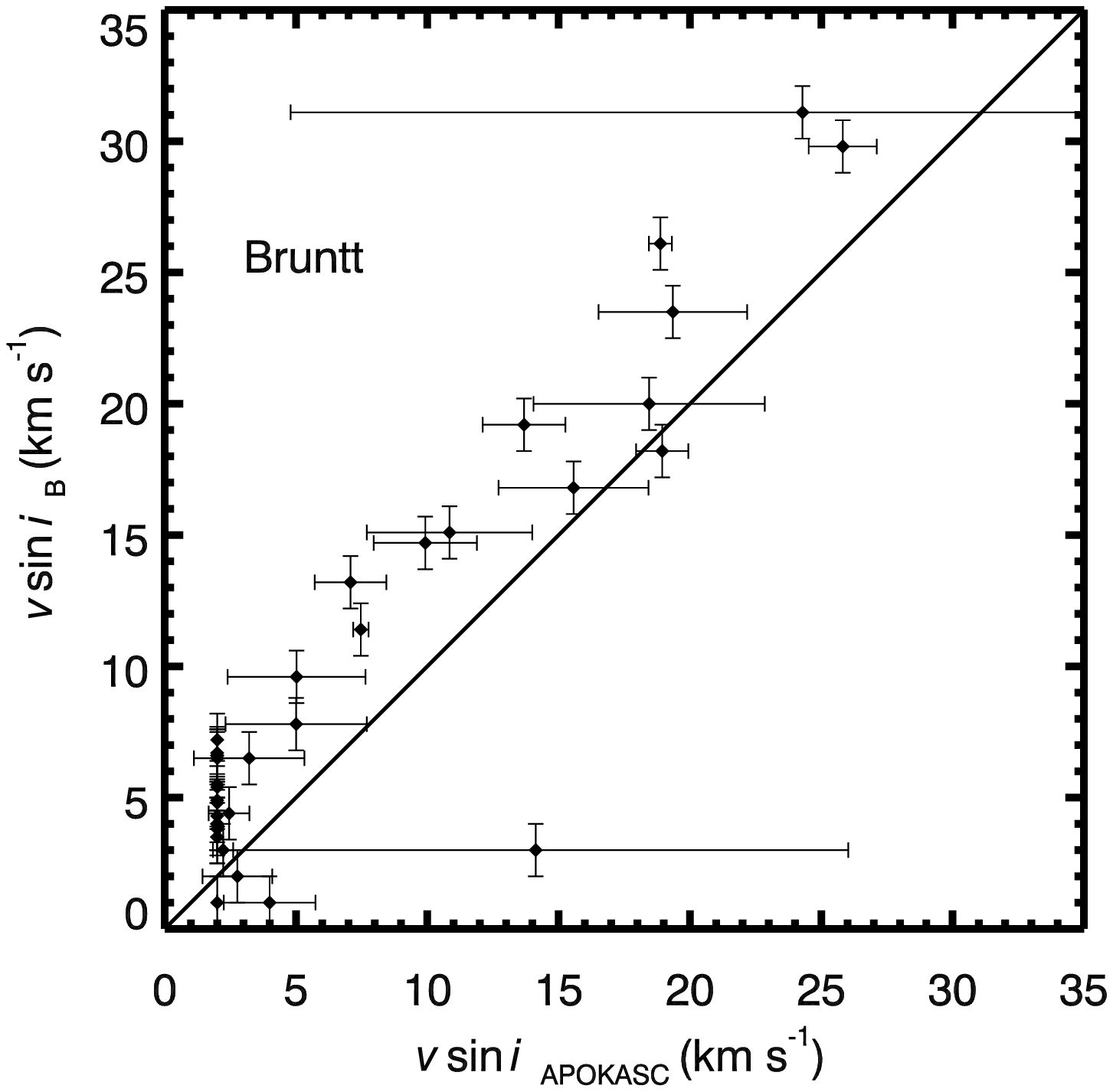}}
\subfigure{\includegraphics[width=.32\textwidth, clip=true, trim=0.5in 1.5in 1.5in 1.6in]{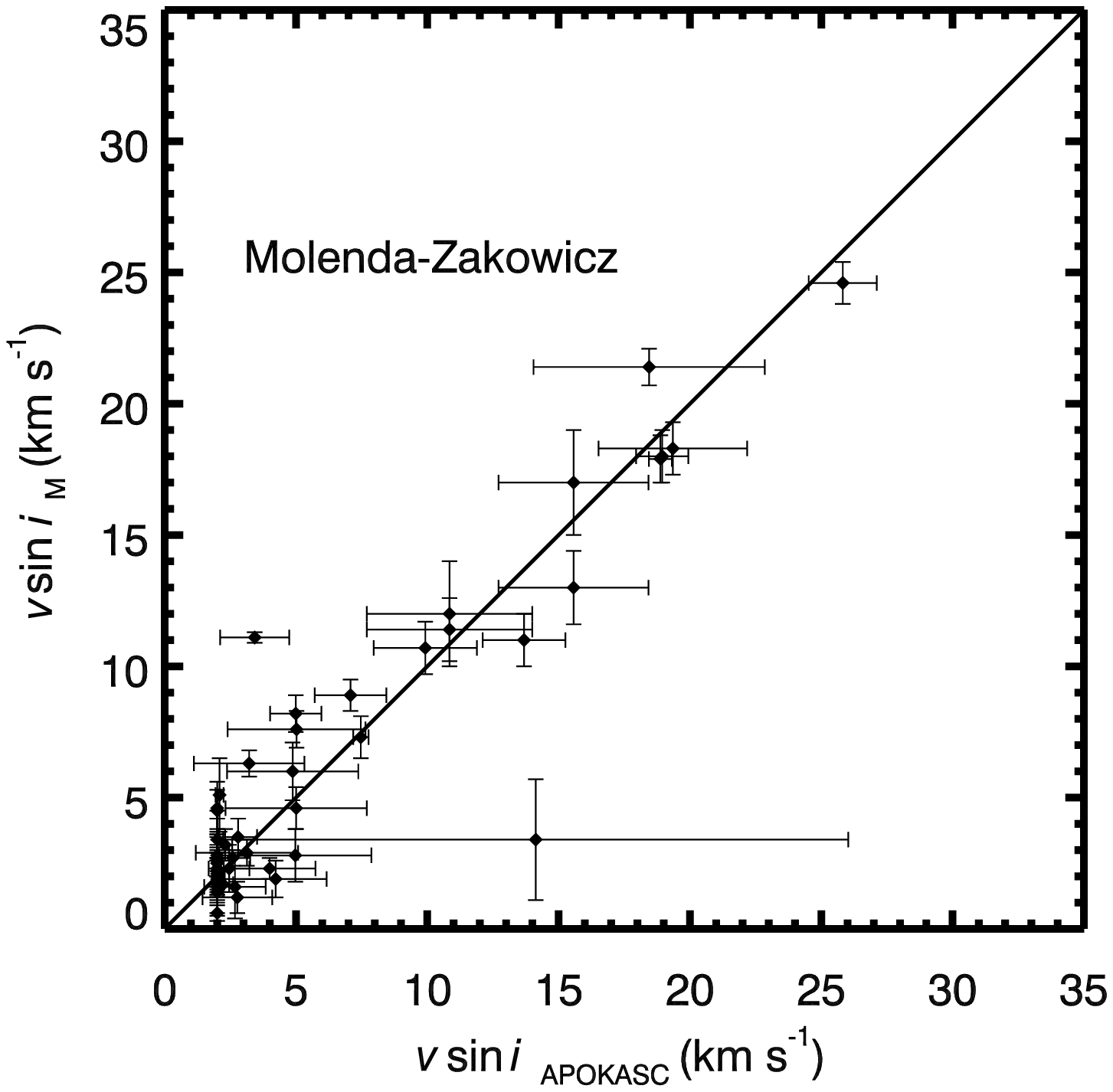}}
\subfigure{\includegraphics[width=.32\textwidth, clip=true, trim=0.5in 0.6in 1.5in 1.6in]{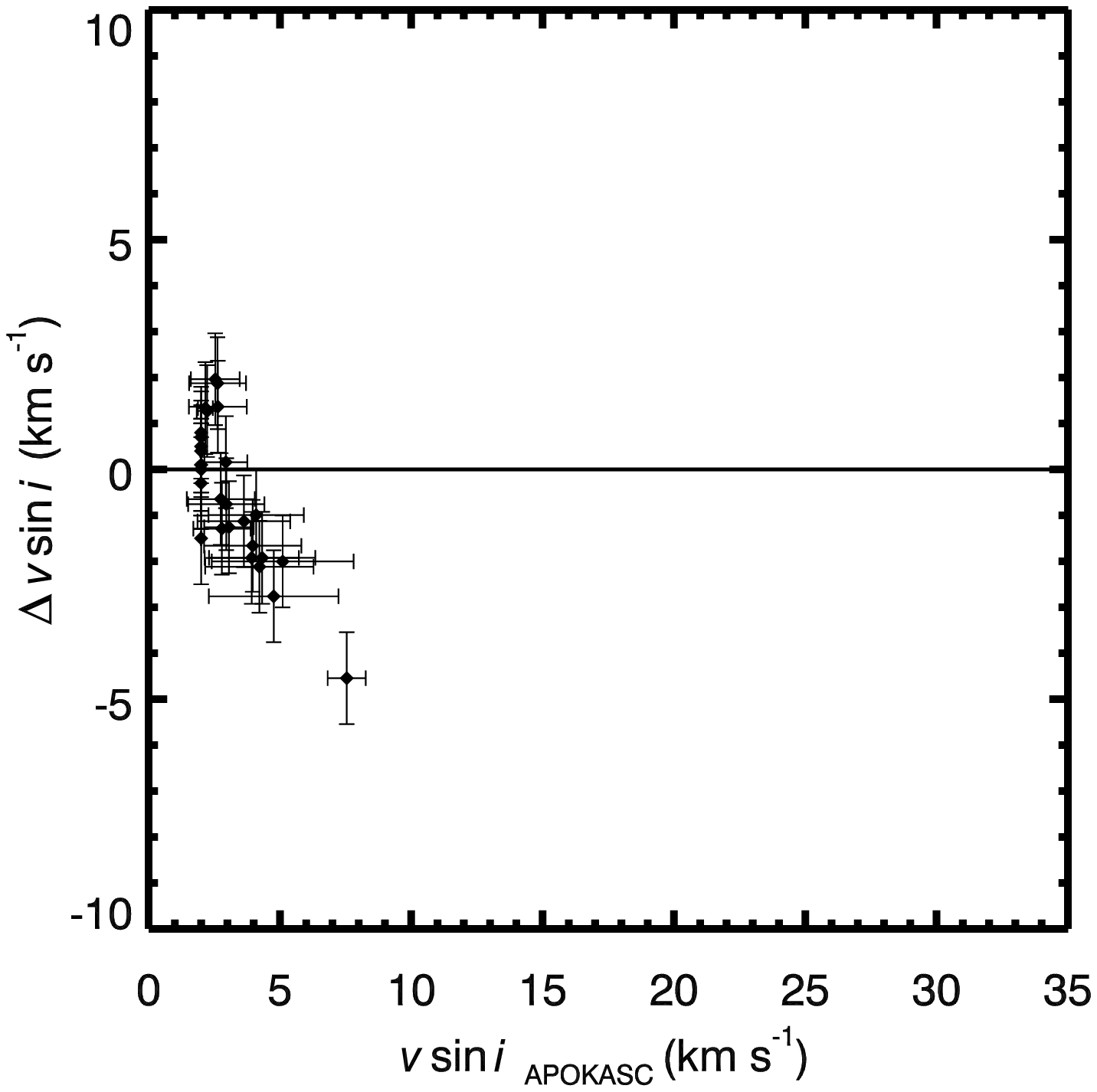}}
\subfigure{\includegraphics[width=.32\textwidth, clip=true, trim=0.5in 0.6in 1.5in 1.6in]{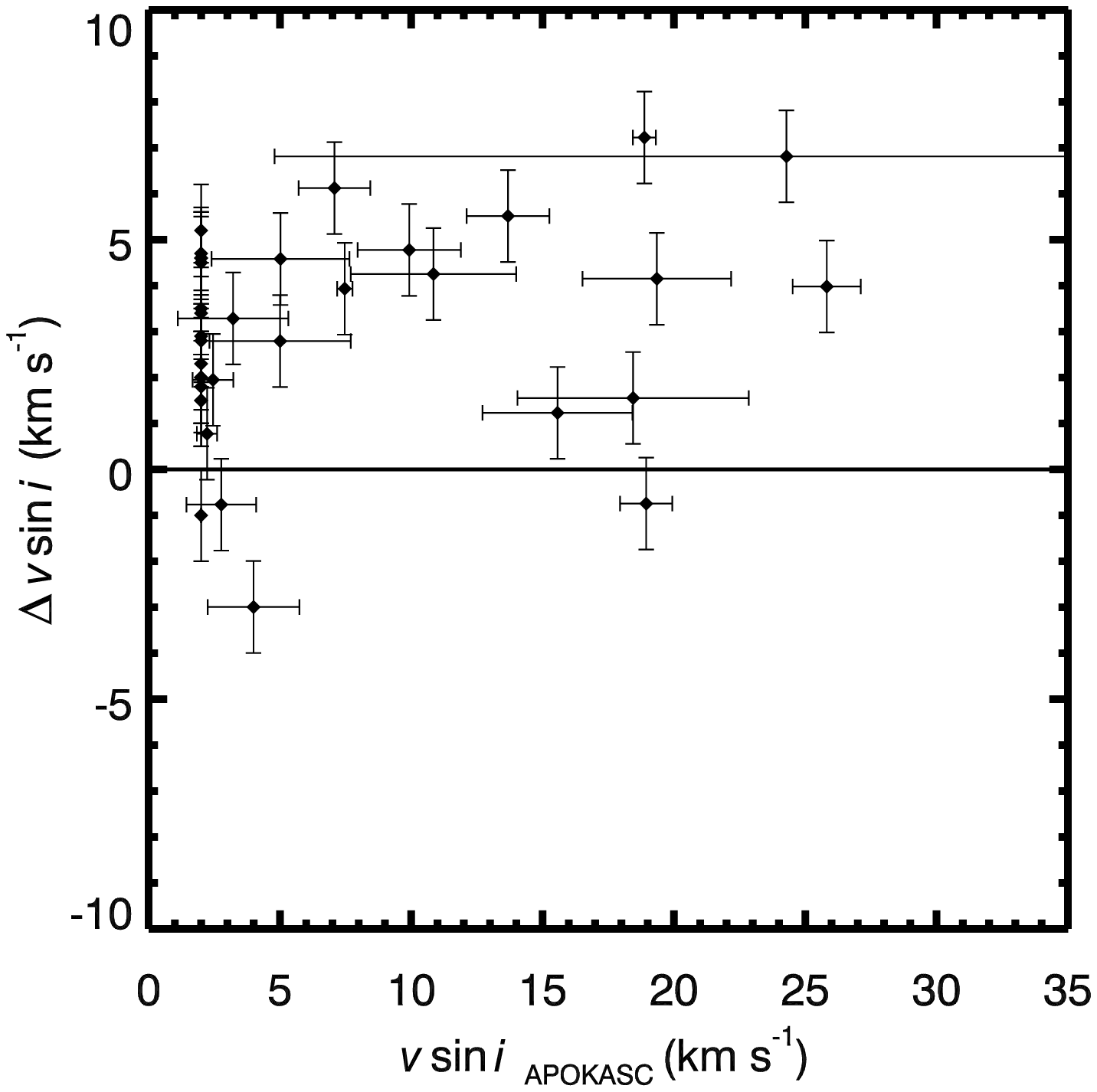}}
\subfigure{\includegraphics[width=.32\textwidth, clip=true, trim=0.5in 0.6in 1.5in 1.6in]{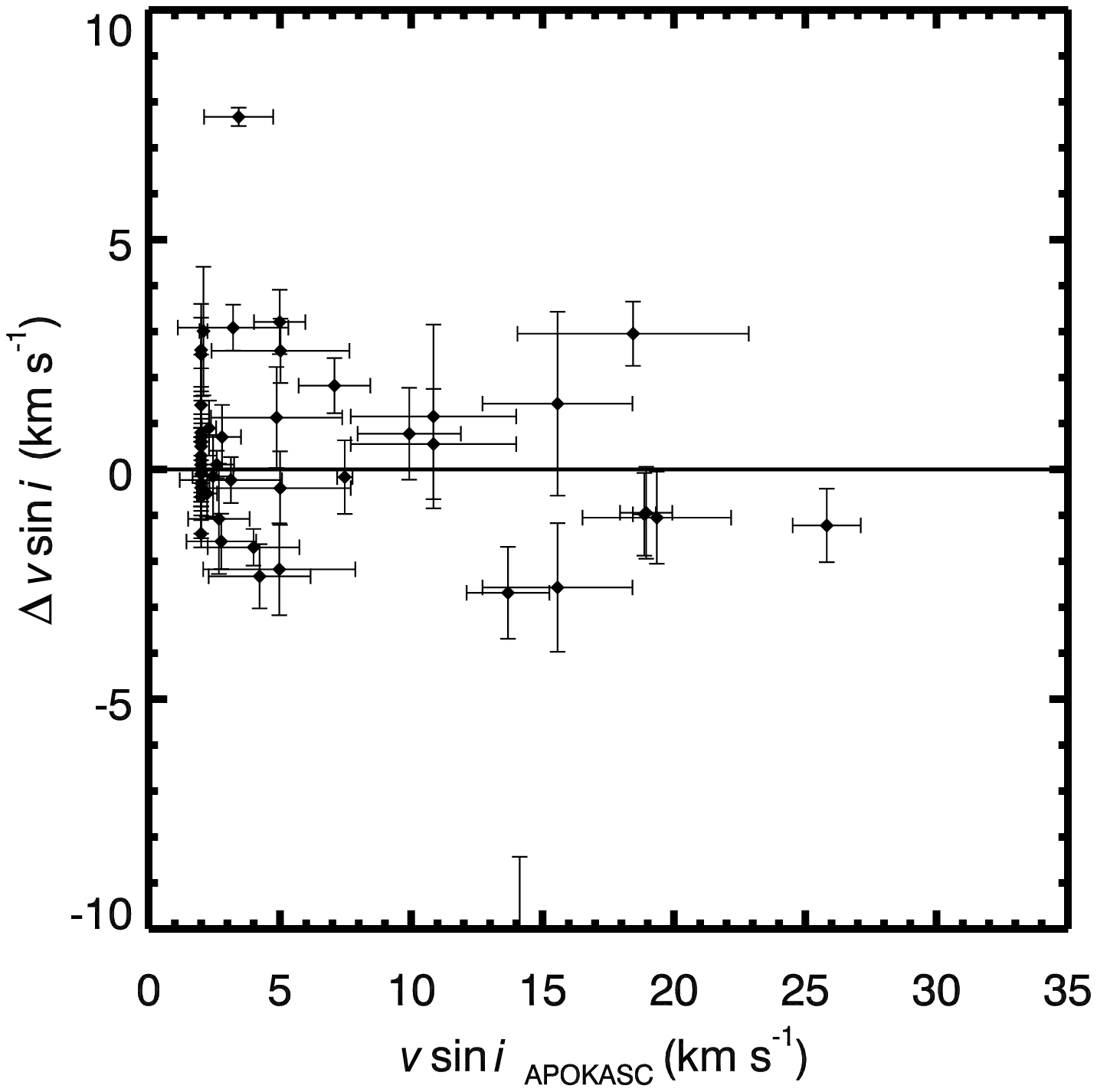}}
\caption{A comparison between the $v\sin i$ values measured in this work and those determined using high resolution spectra. The panels (from left to right) show comparisons with results from \citet{Thygesen2012},  \citet{Bruntt2012}, and \citet{Molenda-Zakowicz2011,Molenda-Zakowicz2013} respectively. Our values above 5 km~s$^{-1}$ are consistent with the Molenda-{\.Z}akowicz results but systematically lower than the Bruntt values.}
\end{figure}
\begin{sidewaystable}[htbp]
 \begin{adjustwidth}{-5cm}{}

\begin{tabular}{rlrrrrrrrrrrlrl}
\hline \hline
\multicolumn{1}{l}{KIC ID} & 2MASS ID & \multicolumn{1}{l}{M } & \multicolumn{1}{l}{$\sigma_M$} & \multicolumn{1}{l}{R} & \multicolumn{1}{l}{$\sigma_R$} & \multicolumn{1}{l}{log \it{g}} & \multicolumn{1}{l}{$\sigma_{log \it{g}}$} & \multicolumn{1}{l}{T$_{\rm eff}$} & \multicolumn{1}{l}{$\rm \sigma_{T eff}$} & \multicolumn{1}{l}{[Fe/H]} & \multicolumn{1}{l}{$\sigma_{[Fe/H]}$} & \multicolumn{1}{l}{$v\sin i$} & \multicolumn{1}{l}{$\sigma_{v\sin i}$} & Evolutionary\\ 
 & & (M$_\sun$)&(M$_\sun$)&(R$_\sun$) &(R$_\sun$) & (cgs) & (cgs) & (K) & (K) & & & (km~s$^{-1}$) &km~s$^{-1}$ & State \\
\hline 

10907196 & J18583782+4822494 & 1.50 & 0.13 & 10.88 & 0.72 & 2.54 & 0.011 & 4786.0 & 86.9 & -0.081 & 0.058 & $<$ 5 & 0 & CLUMP \\ 
10962775 & J18582020+4824064 & 1.21 & 0.13 & 10.94 & 0.45 & 2.44 & 0.011 & 4783.0 & 93.7 & -0.288 & 0.064 & $<$ 5 & 0 & UNKNOWN \\ 
11177749 & J18571019+4848067 & 1.08 & 0.13 & 10.54 & 0.46 & 2.43 & 0.011 & 4703.7 & 82.6 & 0.064 & 0.054 & $<$ 5 & 0 & UNKNOWN \\ 
11231549 & J18584464+4857075 & 1.55 & 0.17 & 13.46 & 0.58 & 2.37 & 0.011 & 4563.6 & 86.3 & -0.031 & 0.057 & $<$ 5 & 0 & UNKNOWN \\ 
11284798 & J18582108+4901359 & 1.29 & 0.16 & 22.26 & 1.07 & 1.85 & 0.012 & 4239.3 & 85.3 & 0.035 & 0.056 & $<$ 5 & 0 & UNKNOWN \\ 
11337883 & J18583500+4906208 & 1.53 & 0.15 & 5.92 & 0.22 & 3.08 & 0.011 & 4837.1 & 84.9 & -0.029 & 0.056 & $<$ 5 & 0 & RGB \\ 
11178396 & J18590205+4853311 & 0.86 & 0.17 & 10.19 & 0.36 & 2.36 & 0.021 & 4854.9 & 111.1 & -0.815 & 0.080 & $<$ 5 & 0 & UNKNOWN \\ 
11284760 & J18581445+4901055 & 1.14 & 0.12 & 10.72 & 0.45 & 2.43 & 0.012 & 4718.5 & 84.1 & 0.015 & 0.055 & $<$ 5 & 0 & UNKNOWN \\ 
11072470 & J19010271+4837597 & 1.10 & 0.17 & 10.90 & 0.51 & 2.40 & 0.018 & 4586.3 & 77.7 & 0.229 & 0.057 & $<$ 5 & 0 & CLUMP \\ 
11072334 & J19004144+4836005 & 1.61 & 0.50 & 8.41 & 1.07 & 2.79 & 0.018 & 4766.5 & 79.2 & 0.158 & 0.051 & $<$ 5 & 0 & UNKNOWN \\

\end{tabular}
\end{adjustwidth}

\caption{Basic properties and projected rotational broadening ($v\sin i$) measured for each star.  Three measurements are made of each rotation velocity and a mean and standard deviation of these measurements is computed. Values below 5 km~s$^{-1}$ are considered nondetections. {Evolutionary states are taken from \citet{Stello2013}.} The full catalog is available online.}

\end{sidewaystable}
Another way to verify our results is to compare the period range allowed by our measurements of $v\sin i$ to actual rotation periods measured from photometric variability. If a star is
magnetically active, its surface will have starspots which will cause periodic brightness modulations as they move across
the star's surface. 
For most giant stars, low activity levels combined with relatively
long rotation periods prevent detection with this technique.
However, as rotation is correlated with activity \citep{Pizzolato2003, Messina2003}, it is possible to derive surface rotation periods for many of the fastest rotating
stars in our sample.

To derive rotation periods, we use the Simple Aperture Photometry
(SAP) time series \citep{Thompson2013} and correct outliers,
jumps, and drifts following the procedures described in \citet{Garcia2011}. This creates what are usually denoted as KADACS (Kepler Asteroseismic Data
Analysis and Calibration Software) light curves. We then follow the
methodology described in \citet{Garcia2014}, using both a
wavelet decomposition \citep{Mathur2010} and the
autocorrelation of the light curve \citep[as in][]{McQuillan2013}. We {then compare the results of these two methods. If the results agree within 10\%, we return the common value as the rotation period. This method is known to derive robust rotation periods { \citep[see][]{Aigrain2015}}.} Each light curve is then visually inspected for agreement with the automatic detection.

While measurement of the rotation periods for the full
sample is postponed to a future paper \citep[for early results see][]{Ceillier2014},
rotation periods are presented in Table~2 for some of the most rapidly
rotating stars in our sample (our rapid and anomalous rotators, see Section 4 for the sample selection procedure). Figure~3 presents a
comparison between the detected rotation period and the maximum rotation period derived from the value
of $v\sin i$ and the seismic radius { (P$_{\rm spec}=\frac{2 \pi \rm R}{v\sin i}$)}. The measured rotation periods all lie at or below the maximum rotation period allowed by our $v\sin i$ measurements, when uncertainties in both measurements are considered. This result confirms that
these stars do indeed have rotation that is peculiar compared to the rest of
the sample.
\begin{table}[htbp]
 \begin{adjustwidth}{-2cm}{}
\begin{tabular}{rllrrrrrl}
\hline\hline
\multicolumn{1}{l}{KIC ID} & 2Mass ID & Type & \multicolumn{1}{l}{$v\sin i$} & \multicolumn{1}{l}{$\sigma_{v\sin i}$} & \multicolumn{1}{l}{P$_{rot}$} & \multicolumn{1}{l}{$\sigma_{Prot}$} & \multicolumn{1}{l}{Mass}& \multicolumn{1}{l}{Evolutionary} \\ 
 & & &(km~s$^{-1}$) &(km~s$^{-1}$)&(days)&(days) &(M$_\sun$)& State\\\hline

3098716 & J19044513+3817311 & Anomalous & 6.94 & 0.73 & -9999 & 0 & 0.86 & CLUMP \\ 
3937217 & J19031206+3903066 & Anomalous & 9.08 & 1.42 & 54.83 & 4.17 & 1.03 & UNKNOWN \\ 
4637793 & J19035057+3946161 & Anomalous & 5.97 & 0.93 & -9999 & 0 & 1.32 & RGB \\ 
4937056 & J19411631+4005508 & Anomalous & 7.27 & 0.94 & 83.69 & 7.63 & 1.66 & UNKNOWN \\ 
5774861 & J19043344+4103026 & Anomalous & 8.81 & 2.14 & 55.6 & 4.53 & 1.17 & CLUMP \\ 
6501237 & J18543598+4155476 & Anomalous & 6.89 & 0.63 & -9999 & 0 & 1.43 & RGB \\ 
8479182 & J18564010+4430158 & Anomalous & 9.25 & 2.13 & -9999 & 0 & 1.25 & CLUMP \\ 
9390558 & J18592488+4556131 & Anomalous & 7.03 & 1.93 & 67.04 & 8.37 & 1.36 & UNKNOWN \\ 
9469165 & J19341437+4605574 & Anomalous & 8.90 & 1.50 & 43.62 & 3.66 & 0.82 & CLUMP \\ 
10128629 & J19053778+4708331 & Anomalous & 7.85 & 1.30 & 79.72 & 7.61 & 1.56 & CLUMP \\ 
10198347 & J19102813+4716385 & Anomalous & 6.94 & 0.82 & 88.14 & 2.44 & 1.37 & UNKNOWN \\ 
11129153 & J19095361+4846325 & Anomalous & 9.31 & 1.46 & 49.07 & 4.47 & 0.88 & UNKNOWN \\ 
11289128 & J19095233+4901406 & Anomalous & 9.72 & 1.22 & 37.71 & 3.17 & 0.87 & UNKNOWN \\ 
11775041 & J19491544+4959530 & Anomalous & 8.85 & 1.17 & 54.08 & 4.74 & 1.18 & CLUMP \\ 
12367827 & J19461996+5107396 & Anomalous & 7.62 & 1.69 & 74.9 & 7.19 & 1.53 & CLUMP \\ 
2285032 & J19063516+3739380 & Rapid & 21.63 & 1.86 & 40.14 & 2.9 & -9999 & UNKNOWN \\ 
2305930 & J19282563+3741232 & Rapid & 13.09 & 0.88 & 33.75 & 2.5 & 0.87 & CLUMP \\ 
3955867 & J19274322+3904194 & Rapid & 12.68 & 1.70 & 32.83 & 2.24 & -9999 & UNKNOWN \\ 
4473933 & J19363898+3933105 & Rapid & 13.60 & 1.13 & 68.45 & 5.51 & -9999 & UNKNOWN \\ 
5193386 & J19343842+4021511 & Rapid & 10.28 & 1.80 & 25.58 & 1.91 & -9999 & UNKNOWN \\ 
10293335 & J19533348+4722375 & Rapid & 12.66 & 1.64 & 49.07 & 3.44 & 2.07 & RGB/AGB \\ 
10417308 & J19460712+4730532 & Rapid & 11.22 & 1.15 & 39.59 & 5.31 & 1.09 & UNKNOWN \\ 
11497421 & J19044946+4929242 & Rapid & 10.19 & 1.65 & 39.86 & 3.22 & 1.11 & UNKNOWN \\ 
11597759 & J18554535+4938325 & Rapid & 11.24 & 1.07 & 46.43 & 3.95 & 0.91 & UNKNOWN \\ 
12003253 & J19015178+5024593 & Rapid & 10.30 & 0.83 & 54.08 & 4.51 & 1.16 & UNKNOWN \\ 

 \hline
\end{tabular}
\end{adjustwidth}
\caption{The full list of rapid and anomalous rotators (see Section 4), along with their projected rotation velocities and their rotation periods. Values of -9999 represent nondetections. {Evolutionary states in this table come from a post-hoc analysis by B. Mosser and many of them were later published in \citet{Mosser2014}.}}

\end{table}

\begin{figure}[H]
 \centering
\subfigure{\includegraphics[width=.45\textwidth, clip=true, trim=.5in .1in .5in 0.5in]{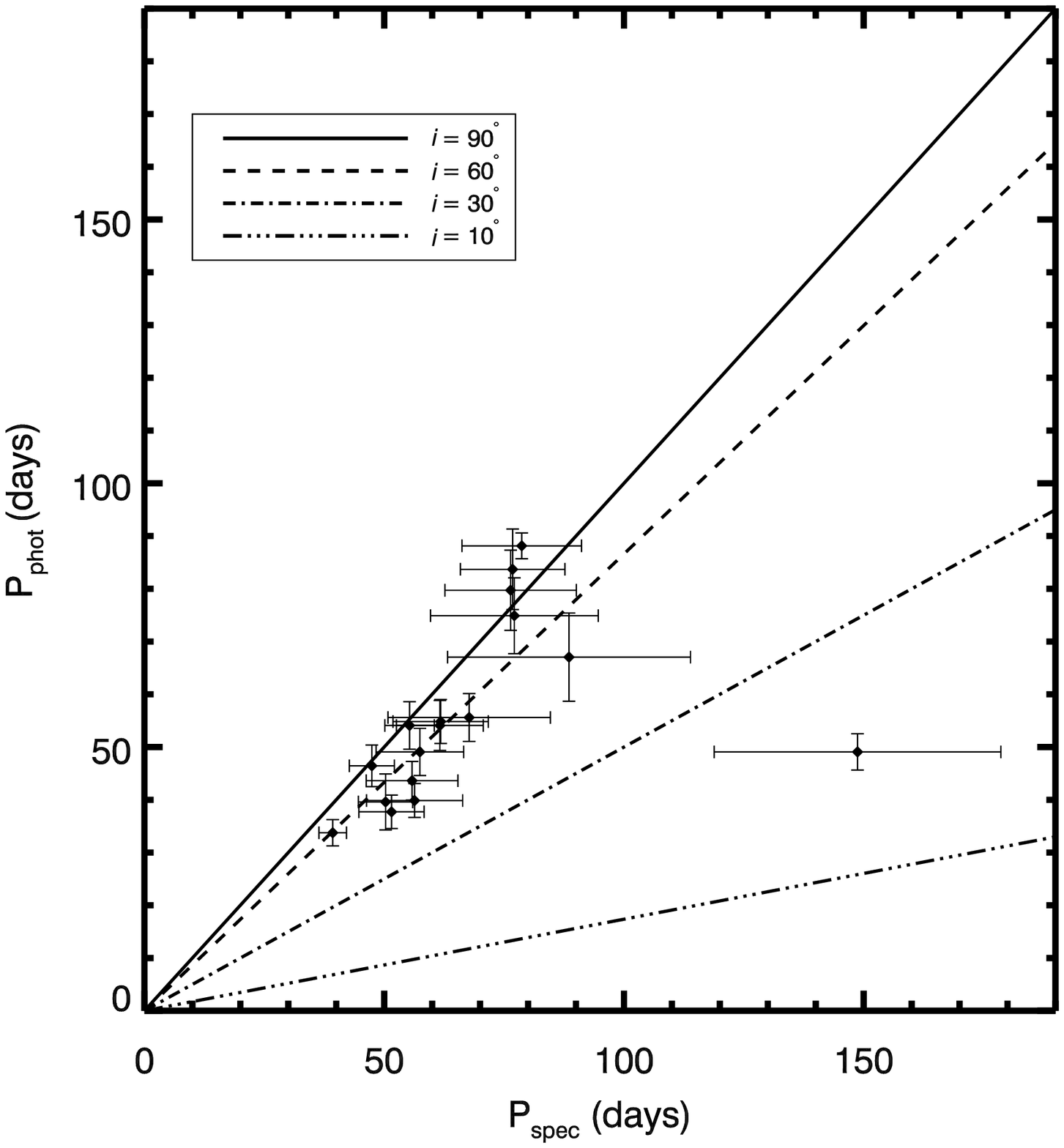}}
\subfigure{\includegraphics[width=.45\textwidth, clip=true, trim=.5in .1in .5in 0.5in]{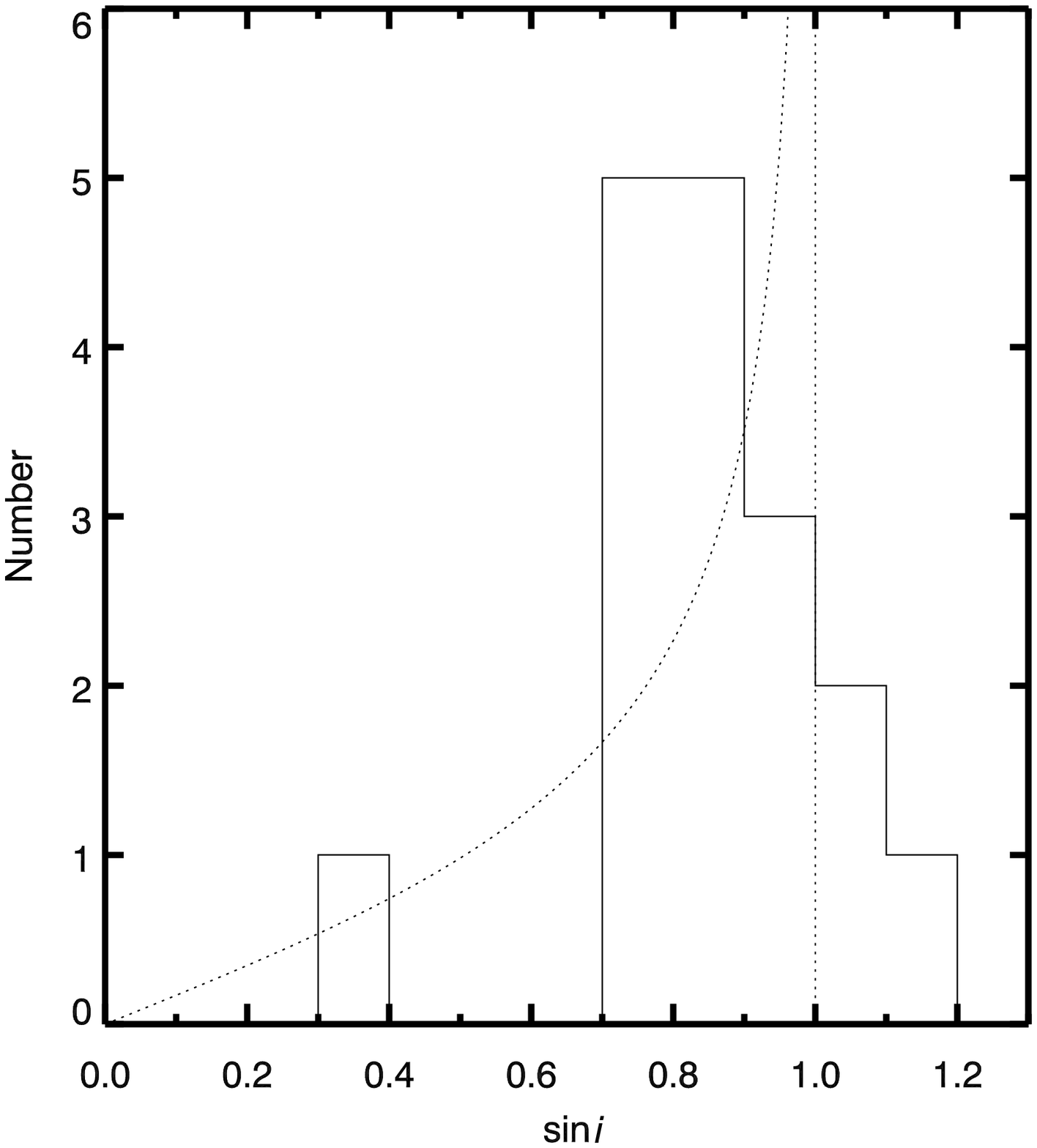}}
\caption{The left panel compares the measured rotation period (P$_{\rm phot}$) to the maximal rotation period computed using the rotation velocity and the stellar radius (P$_{\rm spec}$) for our rapid and anomalous rotators (see section 4). Lines indicate the expected location on the diagram of stars inclined at 90, 60, 30, and 10 degrees. The right {panel} shows the distribution of inclination angles for the same stars computed by comparing the photometric and spectroscopic rotation periods. Dashed lines indicate the expected result for a {random distribution of rotation axes}. }
\end{figure}

\section{Modeling Expected Rotation Rates}
Main sequence stellar rotation rates are mass dependent. Single stars below M $<$ 1.3M$_{\sun}$ (T$_{\rm eff}<$ 6250 K)  rotate slowly \citep{Kraft1970}. Such stars reach the main sequence with a range of rotation periods from 0.1 to 20 days \citep[e.g.][]{Moraux2013, Hartman2010,IrwinBouvier2009}. However, these stars have thick convective envelopes, so they are able to generate magnetized winds that carry away angular momentum \citep{WeberDavis1967} and spin the stars down to rotation rates less than 5 km~s$^{-1}$ by the end of their main sequence lifetime. As these stars expand on the giant branch, angular momentum conservation dictates that their surface rotation slows even more, to less than 1 km~s$^{-1}$. While the star may be spun up somewhat by the dredge-up of internal angular momentum from the core \citep{SimonDrake1989, Massarotti2008} and by structural changes during the helium flash \citep{SillsPinsonneault2000}, such events are not sufficient to increase the surface rotation of low mass stars above 6 km~s$^{-1}$. Mass loss on the upper red giant branch or during the helium flash would reduce surface rates even further for core helium burning stars \citep{SillsPinsonneault2000}.

Stars between 1.3 and 3 M$_{\sun}$ reach the main sequence with a wide, somewhat mass-dependent range of rotation rates \citep[up to 450 km~s$^{-1}$; see e.g.][]{Gray1982,Finkenzeller1985,Alecian2013}. These stars do not have deep surface convection zones and thus do not lose significant angular momentum on the main sequence to winds. The initial range of rotation rates therefore persists to the end of the main sequence phase. As the radius of the star expands on the giant branch, our models indicate that surface rotation rates of all but the most rapidly rotating stars slow down to about 10 km~s$^{-1}$ or less {by the red giant branch bump}. Post-main-sequence angular momentum loss due to mass loss on the lower half of this mass range can further decrease the surface rotation in such stars \citep{vanSadersPinsonneault2013}. Stars with masses below about 2.2 M$_{\sun}$ undergo a giant branch evolution similar to that of low-mass stars, where the rotation continues to slow on the giant branch, except during dredge-up episodes and the helium flash. Stars above about 2.2 M$_{\sun}$ do undergo the dredge-ups that slightly increase their surface rotation rates, but they smoothly, rather than degenerately, ignite their helium cores \citep{Kippy}. This nondegenerate helium ignition only affects the surface rotation in so far as the radius of the star contracts at the beginning of the helium burning phase. However, it does mean that intermediate-mass core helium burning stars are found in the secondary red clump, which is less luminous than the red clump formed by low-mass core helium burning stars \citep{Girardi1999}. While giant branch rotation rates of intermediate-mass stars are therefore expected to be slower than their main sequence rotation rates, the lack of main sequence angular momentum loss means that some intermediate-mass stars could still be rotating faster than 10 km~s$^{-1}$ on the giant branch. We emphasize that in a sample of stars with known masses, it would be possible to exclude such contaminants and tag only rapid rotators that are likely the result of interactions.

To understand the selection effects of applying a single threshold for rapid rotation, we compute the expected rotation rates for stars between 0.6 and 3.0 M$_\sun$ as they ascend the giant branch. For this purpose, we use the Yale Rotating Evolution Code \citep{Pinsonneault1989, vanSadersPinsonneault2012}.  Angular momentum loss due to magnetized winds is accounted for using a modified \citet{Kawaler1988} loss law \citep[see][]{Sills2000, Krishnamurthi1997}. {Specifically, we assume angular momentum loss {$\frac{dJ}{dt}$ depends on the angular velocity $\omega$ as} $\frac{dJ}{dt}=-K\omega^3(\frac{R}{R_\sun})^{0.5}(\frac{M}{M_\sun})^{-0.5}$ up to a Rossby scaled critical rotation rate { ($\omega_{crit}$) which depends on the convective overturn timescale ($\tau_{star}$) of the model as} $\omega_{crit}= \frac{34}{\tau_{star}}$. For stars above the critical threshold, we assume  $\frac{dJ}{dt}=-K\omega\omega_{crit}^2(\frac{R}{R_\sun})^{0.5}(\frac{M}{M_\sun})^{-0.5}$, with $K=2.73 \times 10^{47}$ s.}
Mass loss is not included in our models as for many of the stars in our sample (lower giant branch stars, quickly evolving massive stars, and low-mass stars which have very low mass loss rates) mass loss is negligible and our results would be unaffected by its inclusion. Additionally, for the small subset of stars {(principally intermediate-mass core helium burning stars)} where mass loss could be important, our red clump rotation rates serve as useful upper limits.  We assume solid-body rotation at all times. This is a reasonable approximation for modeling stellar surface rotation rates, because the moment of inertia of the radiative core in red giants is small \citep{SillsPinsonneault2000} and as a result core-envelope decoupling has only a small impact on surface rotation rates \citep{vanSadersPinsonneault2013}.

For solar-mass stars, the assumption of solid-body rotation on the main sequence is well motivated based on both the solar profile \citep{Schou1998} and measurements of solar-like stars \citep{Nielsen2014}. On the giant branch, the assumption of solid-body rotation is less well motivated, but this assumption has a less than 2\% effect on the predicted surface rotation rate. For low-mass stars, we expect {equatorial} giant branch surface rotational velocities less than 0.3 km~s$^{-1}$ (see {the left-hand panel of} Figure 4). Although these models do not include the red clump, analysis by \citet{SillsPinsonneault2000} indicates that structural changes during and after the helium flash should at most increase the predicted rotation rates by a factor of ten at fixed radius. 

For intermediate-mass stars (1.3-3.0 M$_\sun$), determining the expected rotation rate is significantly more complicated. Such systems do not spin down on the main sequence and thus they reach the giant branch with a range of rotation rates. While a detailed statistical analysis of the expected distribution of surface rotation rates of intermediate mass stars is deferred to another paper, we show in the left panel of Figure 4 a demonstrative track of the expected rotational evolution of a 3.0 M$_{\sun}$ star that rotated in the 95th percentile of measured rotation rates in this mass range \citep[250 km~s$^{-1}$ at the base of the {giant branch};][]{ ZorecRoyer2012}. For such a star, rotation decreases to 5 km~s$^{-1}$ on the upper giant branch, but can be as much as 15 km~s$^{-1}$ in the clump. We stress that given this model's location at the top of the mass range and its rapid main sequence rotation, its rotational evolution should be considered an upper limit on the expected giant branch rotation rates. 

We show in the right panel of Figure 4 the expected rotation rates of {some of the most rapidly rotating} stars in our mass range. {{The zones of expected rotation in this plot come from
}models between 0.6 and 3.0 solar masses in steps of 0.2 M$_\sun$, which have disk locking times of 0.8 Myrs and initial rotation rates of $1.0 \times 10^{47}$ rads~s$^{-1}$. This produces velocities of about 250 km~s$^{-1}$ at the base of the giant branch for intermediate mass stars; these models thus provide a reasonable upper limit to the rotation rates that could result from single star evolution.} We expect very slow rotation for low-mass stars and slow rotation for evolved stars between 1.2 and 1.6 M$_\sun$. Rapid and moderate rotation rates are possible for intermediate-mass stars (M $>$ 2.0M$_\sun$) depending on their radius. Given the clearly mass-dependent expectations for giant branch rotation rates, we emphasize the necessity of stellar mass measurements for accurately identifying the full sample of anomalously rotating stars.

\begin{figure}[h]
 \centering
\subfigure{\includegraphics[width=0.49\textwidth, clip=true, trim= 0.5in 0.2in 0.3in 0.5in]{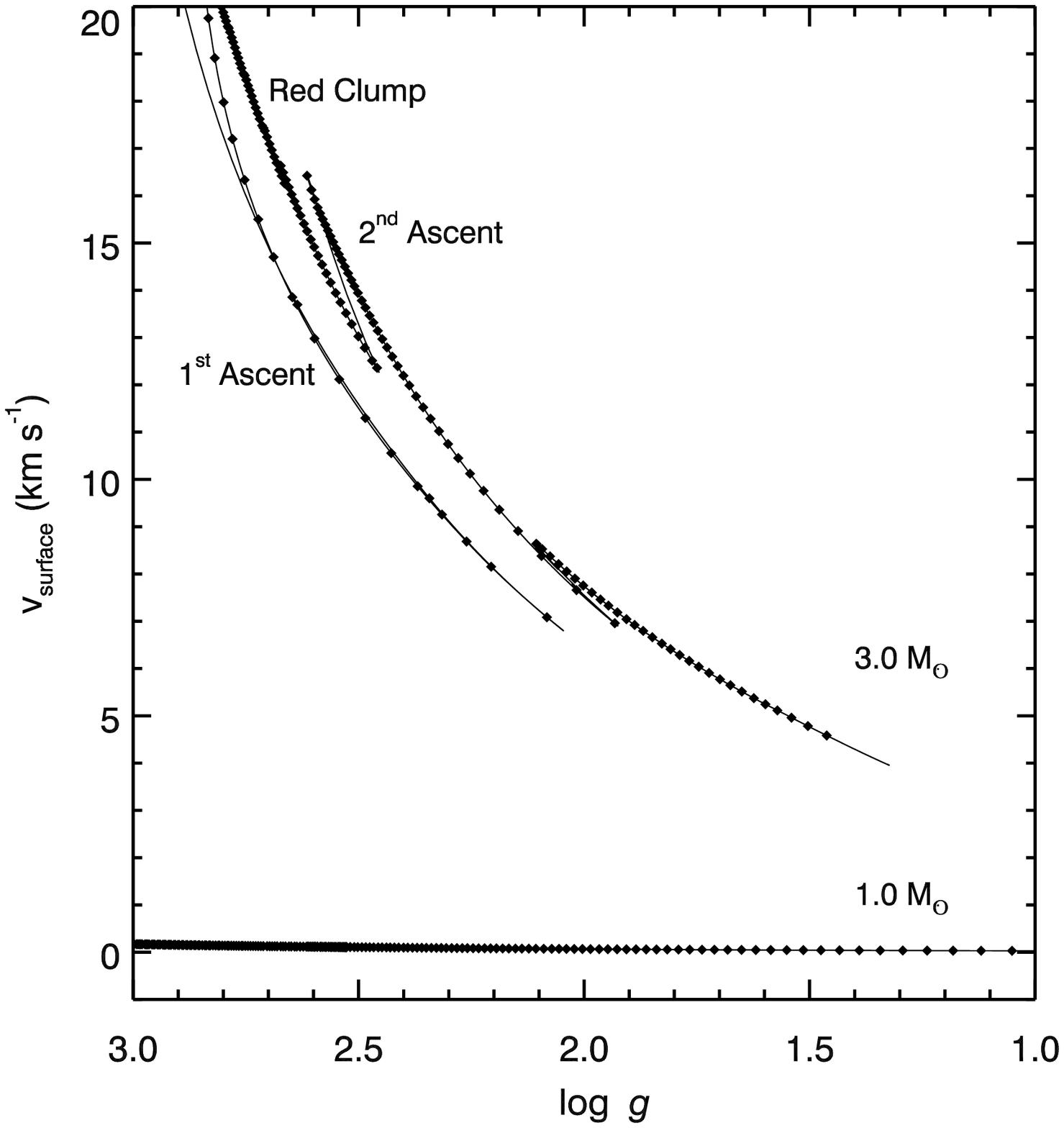}}
\subfigure{\includegraphics[width=0.49\textwidth, clip=true, trim= 0.5in 0.2in 0.3in 0.5in]{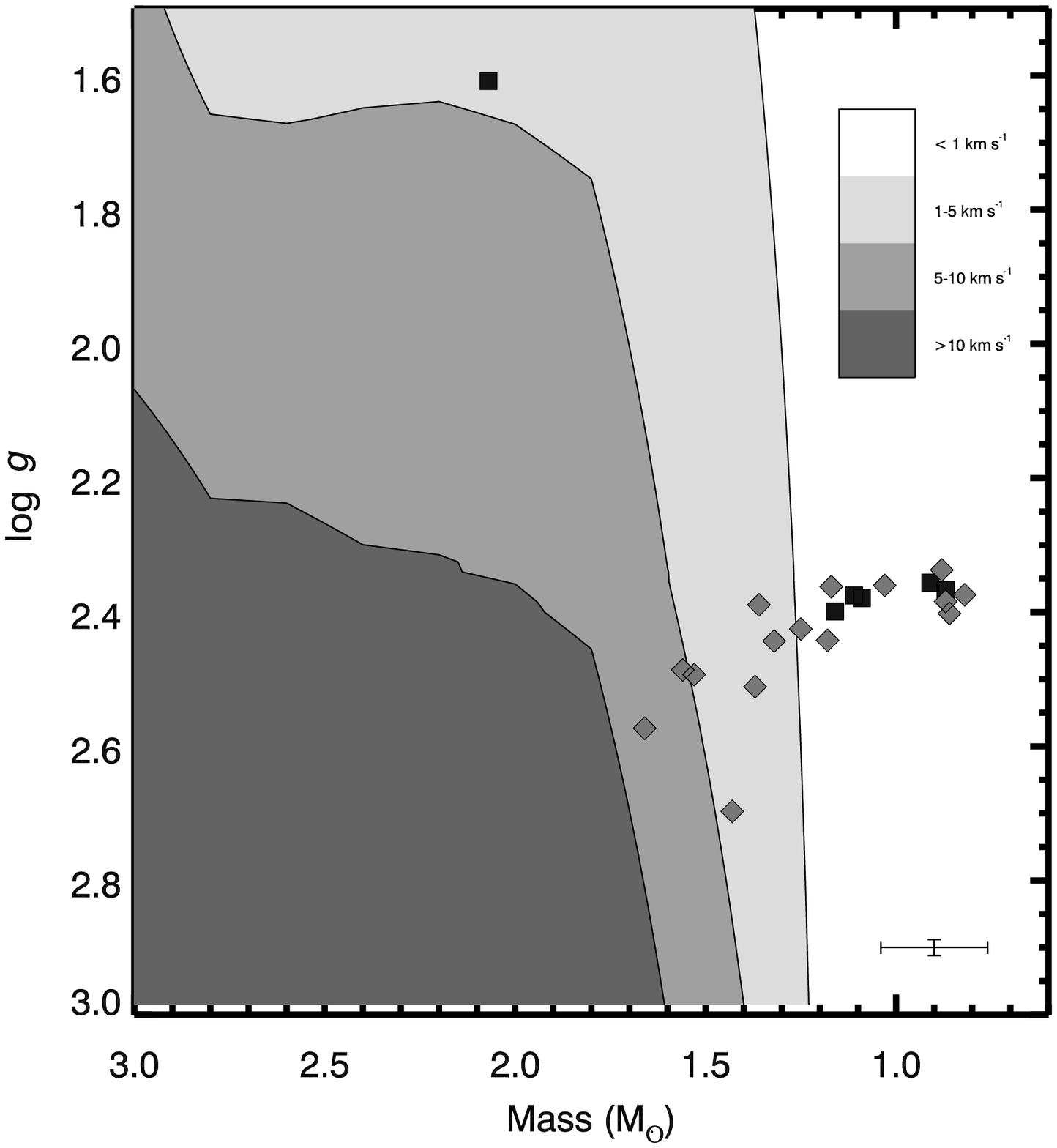}}
\caption{The left panel presents model predictions (solid lines) of the expected post-main-sequence surface rotation rate of both low-mass (1.0M$_\sun$) and intermediate-mass (3.0M$_\sun$) stars. Diamonds indicate steps of 1 million years. As described in the text, the intermediate mass model plotted here should be treated as an upper limit on the expected rotation rates of stars in our sample. In the right panel, we combine many such tracks to identify regimes where we expect rapid, moderate, and undetectable rotation. We also mark for reference the locations of our rapid (black squares) and anomalous (gray diamonds) rotators (see section 4) to demonstrate that this rotation cannot result from single star evolution. Demonstrative error bars are shown in the lower right corner.}
\end{figure}

\section{Rapid and Anomalous Rotator Sample}

The computed projected rotational velocities of the 1950 giants in our sample are listed in Table 1. Any value below 5 km~s$^{-1}$ is listed as a nondetection {(1869 stars)}.
We followed this automated analysis with a visual inspection of targets with automatically computed $v\sin i$ values greater than 10 km~s$^{-1}$ {to determine whether the breadth of the lines were consistent with 10 km~s$^{-1}$ rotation}. This procedure produced a sample of 10 rapidly rotating stars, defined for comparison with previous results by using the standard procedure of identifying all stars with a $v\sin i$ greater than 10 km~s$^{-1}$ as rapid rotators. This threshold is high enough to make false positive detections unlikely.

 We identify all stars rotating between 5 and 10 km~s$^{-1}$ as moderate rotators {(71 stars)}. While less observationally secure, confirmed rotation rates in this range could be the result of an interaction.
Because our sample has measured masses, metallicities, and surface gravities, we also have the ability to identify those moderate rotators that are rotating more quickly than all stars of similar properties as well as the rate that our models predict.  While we believe that the inclusion of such stars will add significantly to our sample size and allow better statistical analysis of stellar interactions, we wish to be conservative in our identification and particularly wish to avoid stars that are scattered into our sample from below our detection threshold and are a measurement-induced tail to the undetectable population (i.e. Eddington bias).

 We therefore focus on selecting the subset of moderate rotators whose rotation is most likely due to a recent interaction, which we refer to as anomalous rotators. To define this sample, we use the mean rotation velocity and the standard deviation computed from the three independent measurements of velocity. We first exclude all moderate rotators whose mean measured rotation rate is not at least one {of its measured} standard deviations above our detection threshold {(43 stars)}. We also do not want to include in our sample any stars whose rotation could be part of the tail of the normal distribution of rotation rates and therefore {iteratively excluded} any star whose rotation is not at least one sigma above that of any other {similar} star not considered an anomalous rotator {(13 stars). See Figure 5 for a visual depiction of this process and our binning into groups of similar stars (within 0.1M$_{\sun}$, within 0.1 dex in $\log g$).} 

We then check that the {artificially broadened template is a good match to the observed spectrum (excludes 2 stars)}, that the anomalous rotation is not the result of a significant ($>$0.3 dex) difference in metallicity from all similar stars  {(excludes 1 star)}, and that the anomalous rotation is inconsistent with predictions by the models {(no stars excluded)}. A list of our 10 rapid and 15 additional anomalous rotators is given in Table 2.

\begin{figure}[h]
 \centering
\subfigure{\includegraphics[width=0.45\textwidth, clip=true, trim=0.8in 0.2in 0.5in 0.3in]{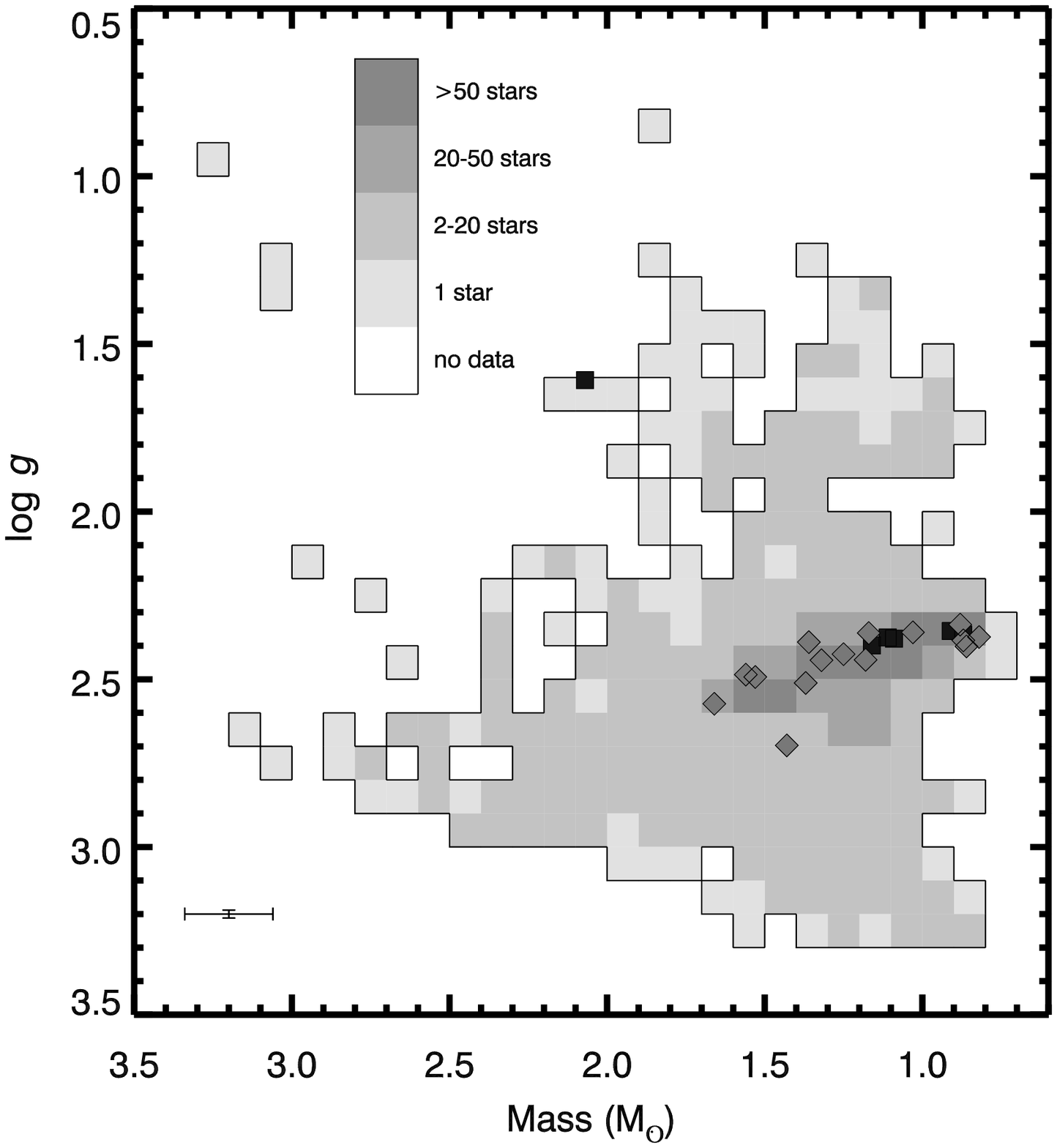}}
\subfigure{\includegraphics[width=0.45\textwidth, clip=true, trim=0.8in 0.2in 0.4in 0.3in]{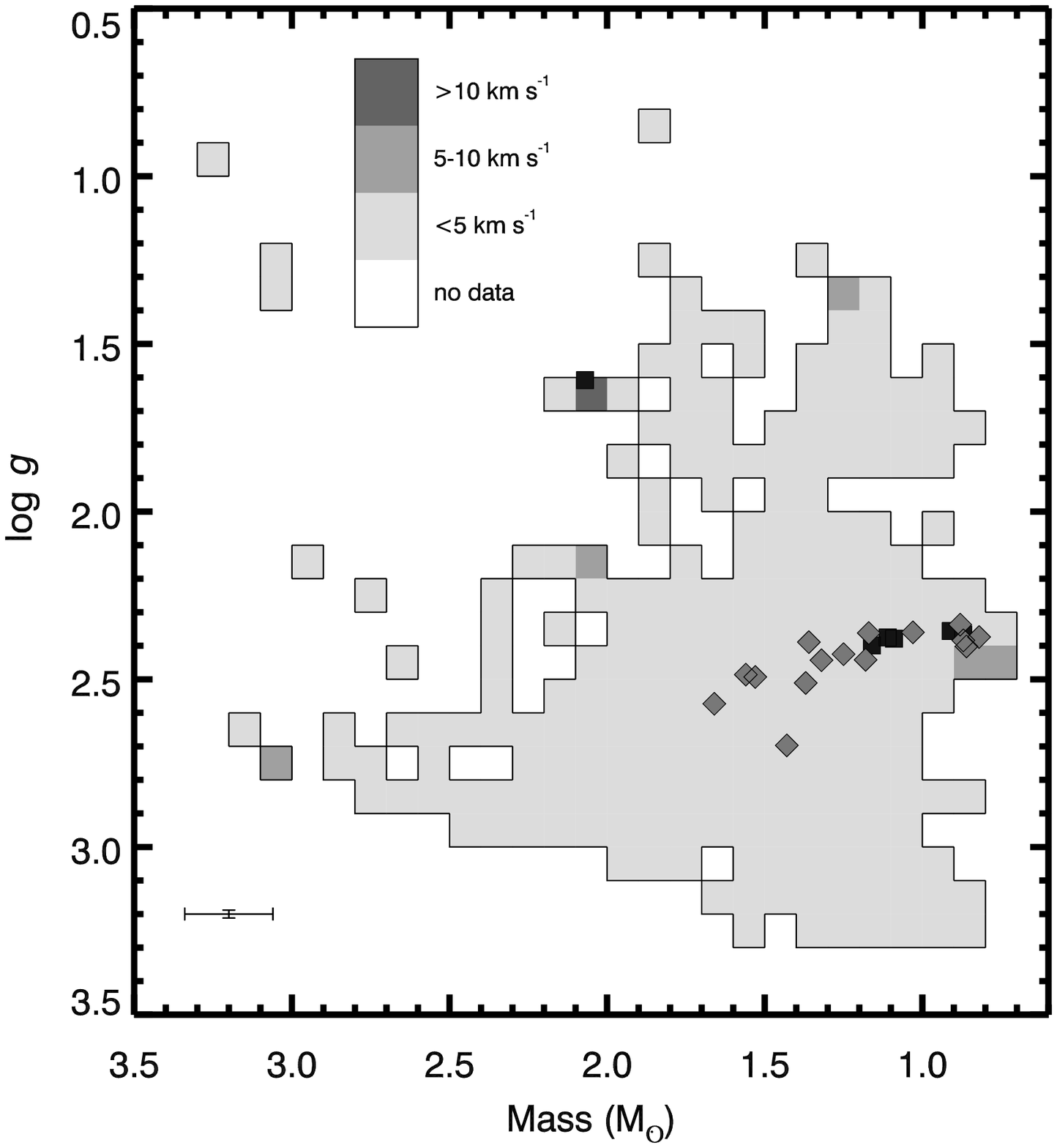}}
\subfigure{\includegraphics[width=0.55\textwidth, clip=true, trim=0.8in 0.2in 0.4in 0.3in]{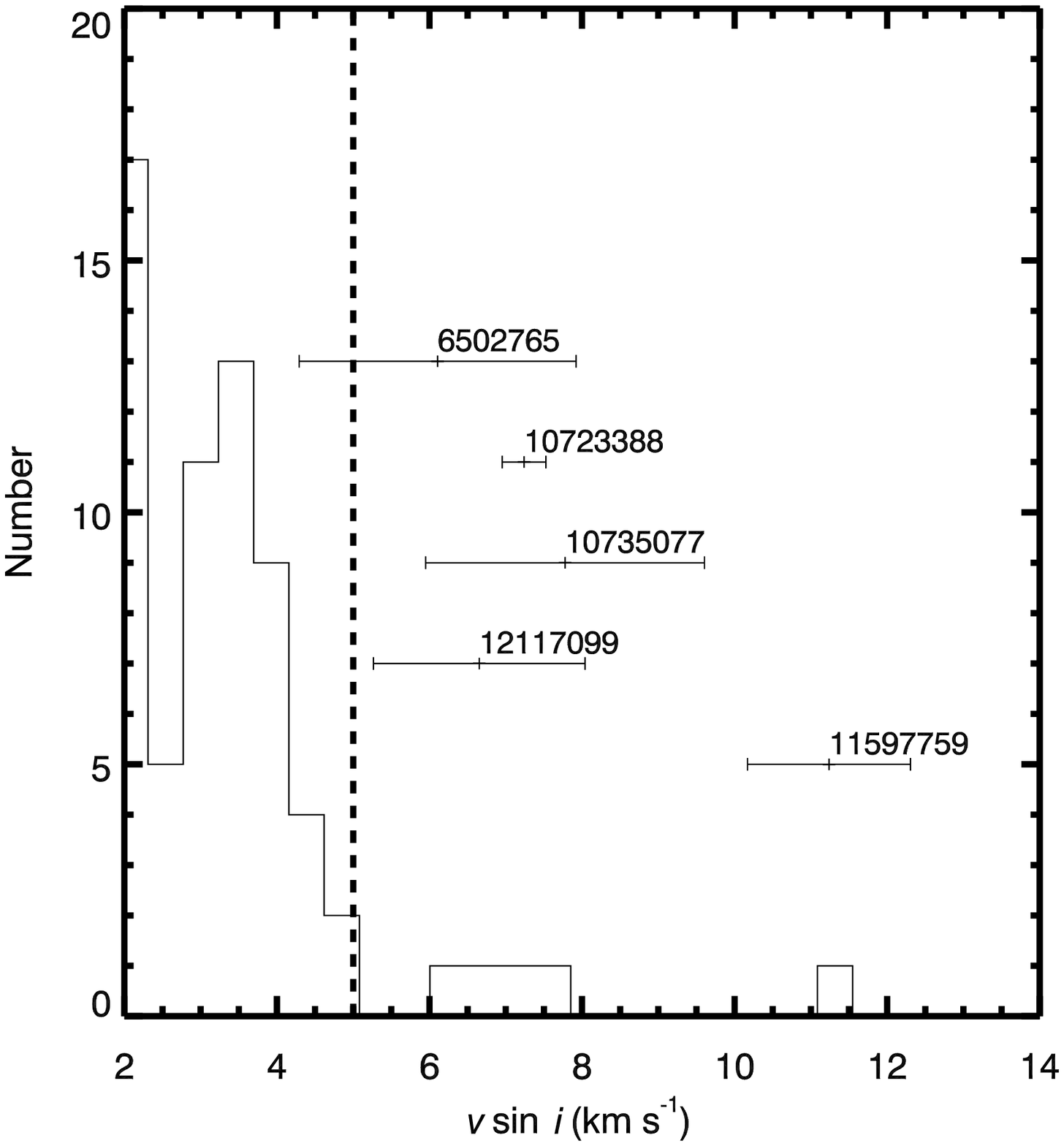}}

\caption{We show the division of stars into bins in mass and surface gravity. In {the top left plot, box color indicates the density of stars in each box; the curved overdense region represents the red clump in this space. In the top right, the box color indicates the average velocity in that box and we emphasize the rapidly and anomalously rotating stars (black squares represent rapid rotators, dark gray diamonds indicate anomalous rotators). The bottom plot shows the distributions of stars within a box.} In the bottom distribution, none of the stars between 6 and 8 km~s$^{-1}$ are considered anomalous because their error bars overlap with either the five kilometer per second measurement floor or a star whose error bars overlap this measurement floor. }
\end{figure}

\subsection{Characterization of Rapid and Anomalous Rotators}
Of our {10} rapidly rotating stars, three are eclipsing binaries whose pulsations are not measured, and {one other lacks a seismic mass measurement} in the APOKASC catalog. This leaves {six} stars whose physical properties, including mass, metallicity and surface gravity, can be compared with the rest of the sample. Because the measured metallicity is strongly affected by the inclusion of rotation in the fit, we use the metallicities measured using the method of Troup et al. (2015{, in prep.}) for our rapid and anomalous rotators, with corrections applied to match published results as in Data Release 12 {\citep{Holtzman2015}. As discussed in Section 2, this tends to decrease the measured rotation velocity by about 1 km~s$^{-1}$, an effect that is not included in our rapid and anomalous rotation selection because this simultaneous fitting is not yet available for the entire APOKASC sample.} 

We aim to use the properties of our rapid rotators to determine the source of the rapid rotation. If {it} results from binary interactions, for example, we would expect more rapid rotators in the low metallicity regime because the close binary fraction is higher in low-metallicity stars \citep{Gao2014}. If stars are most likely to interact at the tip of the red giant branch, where their radii are largest, we would expect rapid rotation to be more common in red clump stars than in red giant branch stars of similar radii. We do indeed find that while less than 4\% of the APOKASC sample is low metallicity ([Fe/H]$<$-0.6, for comparison with Gao et al.), {one of the ten rapid rotators (10\%) is} low metallicity. {While none of our rapid rotators had evolutionary state classifications in \citet{Stello2013}, detailed post hoc analysis (B. Mosser, private communication) indicates that at least one (KIC 2305930) is a clump giant. }

Because the rapidly rotating sample is so small, drawing conclusions about the distribution of interacting stars from this sample is difficult. However, because we have identified a larger sample of stars whose rotation is {likely} the result of interaction, we can increase the sample size by combining our rapidly rotating and our anomalously rotating samples to examine the characteristics of stars which are likely interaction products. There are almost certainly interaction products in our sample which are not included in our list and therefore the fractions we derive should be considered lower limits with complicated selection effects. Nevertheless, this larger group of 21 stars is compared to the distribution of all APOKASC stars to search for trends that could reveal the origin of our rapidly rotating stars. First, we examine evolutionary states. In our sample of 1924 stars with seismic masses and measured rotational velocities, {15}\% are categorized seismically as clump stars, and {10}\% as red giant branch stars. The {rapid and anomalous stars are slightly more likely to be classified as clump giants, although} the small size of the rapidly and anomalously rotating sample of stars with seismic evolutionary state classifications prevents the placement of a strong constraint.  However, if we include uncategorized stars that have clump-like masses and surface gravities (0.7 $<$ M $<$ 2.0 M$_\sun$, 2.3 $<$ log {\it g} $<$ 2.6), where we expect only about 10\% contamination by red giant branch stars \citep{Bovy2014}, we see a 56\% enhancement in the fraction of stars that are anomalously or rapidly rotating {(19 out of 1115 stars in this region versus }{the 12.2 stars we would have expected, a difference significant at the p $<$ 0.05 level)}. 
 This larger sample provides strong support for the idea that stars are most likely to interact at the tip of the red giant branch. We also detect more rapidly rotating stars than expected in the low mass regime, specifically M $<$ 1.2 M$_{\sun}$, although given that these are the stars that are expected to be rotating extremely slowly due to magnetic braking on the main sequence, it is possible that this result is simply an indication that anomalous rotation is easier to identify in this regime and {does not offer information on the} binary fraction as a function of mass.  Finally, rapid and anomalously rotating stars are more likely to have subsolar ({18 out of 25 stars versus 54\% in the full sample) and low ([Fe/H]$<$-0.6) metallicities (3 out of 25 stars versus 3\% in the full sample)}, which is consistent with \citet{Carlberg2011}, and our deductions from the smaller rapidly rotating sample. Because our rapidly rotating stars are most likely to occur in populations where the binary fraction is large and the likelihood of recent interaction is high, our findings are consistent with the idea that rapid and anomalous rotation is the result of a recent interaction. {We show in Figure 7 the locations in mass and metallicity space of the rapid rotators compared to the whole sample}.

\section{Explaining Unexpected Rotation}
{As shown in Figure 4b, we expected a fraction of the intermediate mass stars to be rotating at detectable rates; it is clear from Figure 5 that this population was not detected in our sample. We suggest several explanations for this difference between our models and the data. The first is that a population of moderately and rapidly rotating stars does exist but was randomly excluded from our sample. However, given that our sample contains 71 stars with masses above 2.2 M$_\sun$, we find this to be unlikely. A second option is these faster rotating stars do exist, but were preferentially excluded by our sample selection criteria. This could plausibly occur if, for example, rotation suppresses pulsations, and should be immediately obvious when looking at the full sample of Kepler data. All of our other explanations assume that this expected population of intermediate-mass stars with fast rotating surfaces does not exist. It is possible, although unlikely, that the rotation distribution measured by \citet{ZorecRoyer2012} is biased towards high rotation velocities and thus not an appropriate starting point. Another possibility is that significant post-main-sequence angular momentum loss occurs in intermediate mass stars that is not included in our models. This could come in the form either of enhanced magnetic wind loss on the giant branch or significant mass loss near the tip of the red giant branch which carries away the angular momentum in a thermal wind. Finally, we suggest that radial differential rotation could be concentrating angular momentum into a fast rotating core and leaving behind a slowly rotating surface. We plan to explore each of these possibilities further in an upcoming paper. }

{While we do not see the general increase in rotation rate that we expected at higher masses,} we have identified {25} stars whose rotation indicates a nonstandard evolution history. In this section, we suggest three possible mechanisms for increasing the rotation rates of stars and attempt to categorize our anomalous rotators as members of these groups. The possible causes for rapid rotation that we investigate are interaction with a binary companion, merger with another stellar object, and accretion of material.

\subsection{Binary Interaction}
The work of \citet{Carlberg2011} indicates that the exchange of orbital and spin angular momentum in tidally interacting binaries should, after accounting for inclination effects, cause approximately 2\% of red giants to have $v\sin i$$>$10 km~s$^{-1}$. {Of the 8 stars in our sample that are listed in the Kepler Eclipsing Binary Catalog \citep[][see keplerebs.villanova.edu for the updated list used here]{Slawson2011}, three are rapid rotators. This is significantly higher than the approximately 10\% of binary stars expected to have a companion close enough to cause tidal synchronization on the giant branch \citep{Carlberg2011}, but consistent with the tendency of eclipsing binaries to be close binary systems. Given the binary periods from the eclipsing binary catalog and the spectroscopic surface gravities from APOGEE, we can compute the stellar rotation velocity we would expect if the binary systems are tidally synchronized, edge on, and have a mass equal to that of the average star in our sample (M = 1.36 M$_\sun$) (See Table 3 for details of these 8 systems). This analysis picks out two close (P$<$50 days) binaries whose velocities and rotation periods indicate that they are {likely close to or exactly} tidally synchronized (KIC 3955867 and 5193386), as well as two close binaries which are curiously either not tidally synchronized or significantly misaligned (KIC 3128793 and 4758368). There is also one wide binary (P $\approx$ 103 days) that is rotating much faster than tidal synchronization would predict (KIC 4473933). We suggest that either the orbit is eccentric or that an unseen third body has previously affected this star.}

\begin{table}[htbp]
 \begin{adjustwidth}{-2cm}{}
\begin{tabular}{rlrrrrrrrr}
\hline\hline
\multicolumn{1}{l}{KICID } & 2MASS ID & \multicolumn{1}{l}{log \it{g}} & \multicolumn{1}{l}{$\sigma_{\rm log \it{g}}$} & \multicolumn{1}{l}{R$_{\rm inf}$} & \multicolumn{1}{l}{P$_{\rm binary}$} & \multicolumn{1}{l}{P$_{\rm rot}$} & \multicolumn{1}{l}{$\sigma_{\rm Prot}$} & \multicolumn{1}{l}{Predicted } & Measured \\ 
\multicolumn{1}{l}{} &  & \multicolumn{1}{l}{(spec)} & \multicolumn{1}{l}{(spec)} & \multicolumn{1}{l}{} & \multicolumn{1}{l}{} & \multicolumn{1}{l}{} & \multicolumn{1}{l}{} & \multicolumn{1}{l}{$v\sin i$} & $v\sin i$ \\ 
\multicolumn{1}{l}{} &  & \multicolumn{1}{l}{(cgs)} & \multicolumn{1}{l}{(cgs)} & \multicolumn{1}{l}{(R$_\sun$)} & \multicolumn{1}{l}{(days)} & \multicolumn{1}{l}{(days)} & \multicolumn{1}{l}{(days)} & \multicolumn{1}{l}{(km~s$^{-1}$)} & (km~s$^{-1}$) \\ 
\hline
3128793 & J19364967+3813244 & 2.9 & 0.2 & 6.5 & 24.7 & -9999.0 & -9999.0 & 13.4 & \multicolumn{1}{r}{5.6} \\ 
3955867 & J19274322+3904194 & 2.8 & 0.2 & 7.6 & 33.7 & 32.8 & 2.2 & 11.4 & \multicolumn{1}{r}{12.7} \\ 
4473933 & J19363898+3933105 & 2.8 & 0.2 & 7.7 & 103.6 & 68.5 & 5.5 & 3.7 & \multicolumn{1}{r}{13.6} \\ 
4758368 & J19394473+3951089 & 2.3 & 0.2 & 13.4 & 3.7 & -9999.0 & -9999.0 & 181.2 & $<$ 5 \\ 
5193386 & J19343842+4021511 & 3.2 & 0.2 & 4.6 & 21.4 & 25.6 & 1.9 & 10.9 & \multicolumn{1}{r}{10.3} \\
6757558 & J18574915+4212172 & 2.9 & 0.2 & 7.1 & 421.2 & -9999.0 & -9999.0 & 0.8 & $<$ 5 \\ 
7431665 & J19093039+4300341 & 2.6 & 0.2 & 9.4 & 281.4 & -9999.0 & -9999.0 & 1.7 & $<$ 5 \\ 
9540226 & J19480815+4611544 & 2.3 & 0.2 & 13.4 & 175.5 & -9999.0 & -9999.0 & 3.9 & $<$ 5 \\ 
 \hline
\end{tabular}
 \end{adjustwidth}
\caption{{Details of the eight known eclipsing binary systems in our sample. We list the binary period, the spectroscopic surface gravity, the radius we infer from these measurements (see text), as well as the rotation velocity we would predict. We note that using the seismic radii (available for KIC 6757558 and KIC 9540226) does not significantly alter our predicted rotation velocities. These predictions can be compared to measured spot periods and rotational velocities where available. }}

\end{table}

If we examine all 1924 stars with seismic masses in the APOKASC sample using the Carlberg threshold for rapid rotation ($v\sin i$ $>$ 10km~s$^{-1}$), only {six} stars are rapidly rotating. Even if all of these objects are rapidly rotating due to binary interactions, they comprise only {0.3\%} of the sample { with known masses}, significantly below the predicted occurrence rate. We therefore conclude that the APOKASC sample does not include the same fraction of close binary stars as the field. We suggest that this result could be due either to the suppression of seismic oscillations in close binary systems \citep{Gaulme2013} or more simply due to the fact that the asteroseismic analysis methods are not designed to automatically disentangle the combined oscillations of two different stars, although such reduction may be possible in 1\% of the oscillating sample \citep{Miglio2014}. This conclusion could be more directly tested using radial velocity (RV) measurements of the sample or possibly the time delay methods of \citet{Murphy2014}.

Though we have a lower than expected probable binary fraction, we still attempt to identify whether any of our anomalous rotators are in fact binary systems. We check for RV variability in the {15 out of 25} stars in our sample which have multiple APOGEE observations. The expected RV jitter for giant stars {ranges from 0.03 to 0.5 km~s$^{-1}$ for surface gravities between 3 and 1 \citep{Hekker2008}}; RV variations significantly larger than 1 km~s$^{-1}$ are considered secure detections for this instrument \citep{Deshpande2013}. While our known binary systems have RV variations on the level of 20-100 km~s$^{-1}$, the majority of the unidentified anomalous rotator sample varies at a level of 0.2-0.5 km~s$^{-1}$. Only KIC 10293335, which changes by 1.1 km~s$^{-1}$, has a possibly significant radial velocity variation. Comparison of the template spectra to the individual visit spectra for each star also fails to identify double-lined spectra or strong asymmetries in the absorption lines. This result appears to exclude most of our anomalous rotators from being in either close in or equal mass binary systems.

We do photometrically detect one possible close to equal mass binary, KIC 6501237. While we have only one spectrum of this star and therefore cannot check for radial velocity variability, careful analysis of the light curve indicates two distinct signatures of seismic variability (see Figure 6). The brighter star has a seismic mass of 1.39 M$_{\sun}$ and the dimmer star has a mass of 1.31 M$_{\sun}$; the seismic radii are 8.7 and 5.9 R$_{\sun}$ respectively. Given measurement errors, this would be consistent with two stars of similar age and thus a possible binary. However, inspection of the high-resolution UKIRT image  of this location indicates that there are two point sources separated by about three arcseconds \citep[see][for details of the camera, photometric system, pipeline processing, and science archive, respectively]{Casali2007, Hewett2006, Irwin2008, Hambly2008}. If these objects are the two giants visible in the \textit{Kepler} photometry, it would explain why only one set of lines is visible in the APOGEE spectrum, which is taken with a 2'' diameter fiber. Using the distance of about 1400 parsecs computed by \citet{Rodrigues2014}, we find that if these stars are in fact a binary pair, they are separated by at least 4000 AU. {However, the seismic radii suggest that the contrast in apparent brightness between the two stars should be about 1 magnitude if they are at the same distance. The UKIRT image shows a contrast between these two sources of 2.5 magnitude. We thus conclude that if these two objects are the two pulsating giants that we observe, they are unlikely to be a binary system.} { The two remaining viable explanations for this system are therefore that either two unrelated red giants of similar mass happened to be projected very close together on the sky or that a binary system of two oscillating red giants was observed such that two sets of lines were not distinguishable in the spectrum and a spurious background star is located nearby.}

\begin{figure}[H]
 \centering
\includegraphics[width=0.6\textwidth, angle=90]{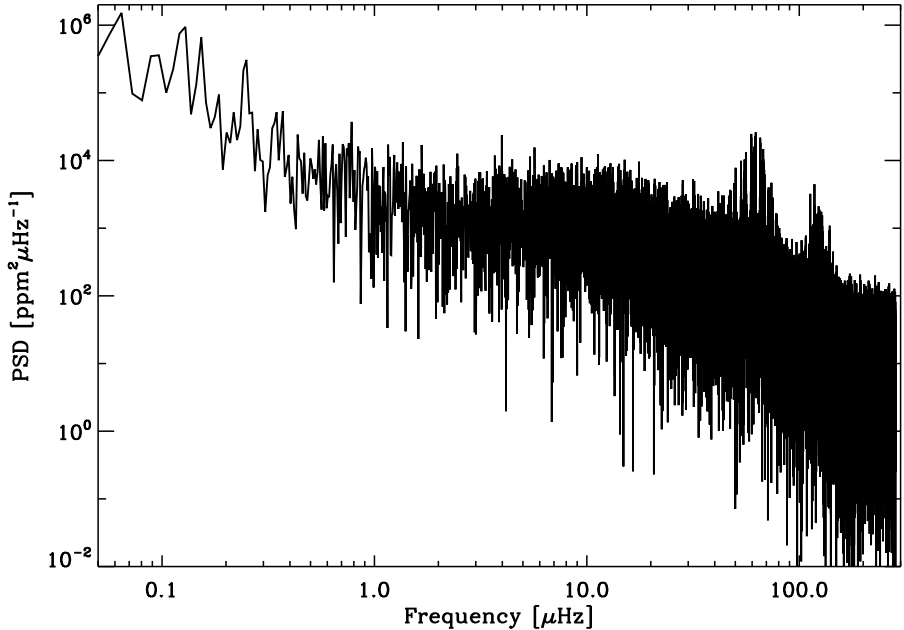}
\caption{The {Lomb-Scargle power density spectrum} of KIC 6501237. Two distinct sets of seismic oscillations are visible, with frequencies of maximum power of 62 and 120 $\mu$Hz. }
\end{figure}

\subsection{Merger Products}

We expect that the descendants of merger products in the APOKASC sample will, like blue stragglers, be less evolved than their mass and metallicity would suggest. The most obvious group of stars of this type will be giants significantly more massive than the sample of stars of similar metallicity, as stars in this metallicity and mass range which are not mergers will have already evolved off the giant branch. Figure 7 presents the distribution of the APOKASC sample in mass-metallicity space and marks both the APOKASC and corrected Troup et al. (2015) metallicities for the rapid and anomalous rotators. We highlight KIC 10293335, an outlier that is almost half a solar mass more massive than all other stars of similar metallicity. Attempting to model this star using either metallicity under the assumption that it is not a merger product produces an age below 1 Gyr using both the YREC and PARSEC \citep{Bressan2012} models. This would make this star at most half as old as any other star of similar metallicity, which suggests that KIC 10293335 is likely to be the result of a merger.

\begin{figure}[h]
 \centering
\includegraphics[width=1.0\textwidth, clip=true, trim= 0.6in 0.2in 0.3in 0.5in ]{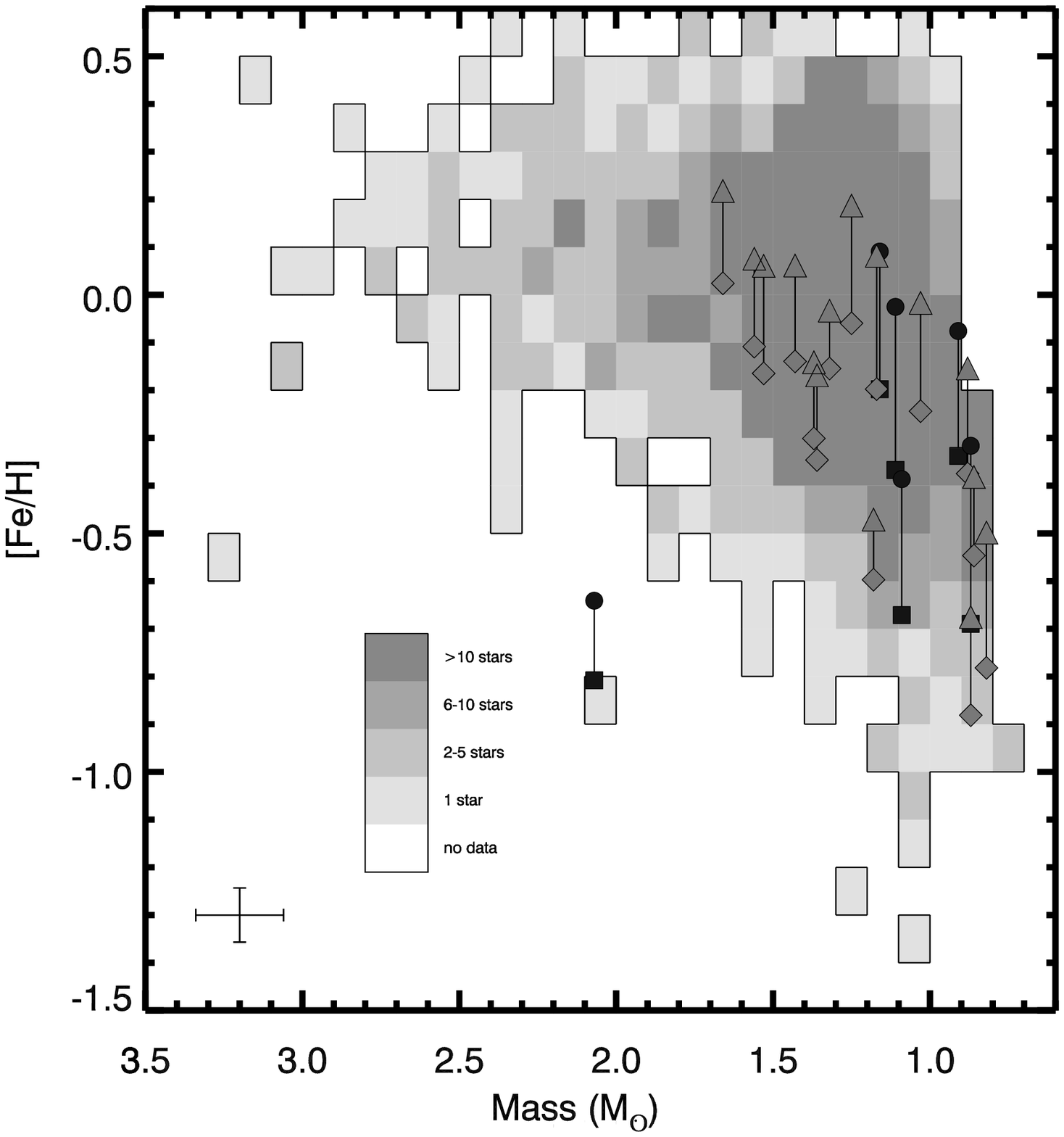}
\caption{This figure displays the number of stars in the sample as a function of mass and metallicity. APOKASC values for rapid and anomalous rotators are marked with black squares and grey diamonds, respectively. These measurements are connected to the results of a simultaneous fitting of the rotation and metallicity (marked with black circles and gray triangles) which tended to increase the measured metallicity by 0.2 dex. We particularly emphasize KIC 10293335 (lower left) as anomalously massive for its metallicity and thus likely the result of a merger. }
\end{figure}

\subsection{Accretion}
Detection of significant numbers of close-in giant planets at rates between 3/1000 stars \citep{Gould2006} and 12/1000 stars \citep{Wright2012} leads to the logical conclusion that at minimum 1\% of stars should accrete mass from planetary size companions as their radii expand on the giant branch. Work by \citet{Carlberg2011} suggests that the accretion of a few Jupiter masses would be sufficient to spin up a solar mass star with a radius ten times that of the sun to a surface rotation rate of 10 km~s$^{-1}$. In addition to the increased rotation rate, we expect that the results of mass accretion might include anomalous lithium abundances and enhanced surface metallicity compared to core metallicity \citep{Carlberg2012}.

{In combination with such measurements, measurements of mixing diagnostics (e.g. carbon isotope ratios or carbon to nitrogen ratios) can help to exclude interpretations of stellar abundance and rotational anomalies that rely on single star evolution. We choose in this work to use the $^{12}$C/$^{13}$C ratios as our mixing indicator as carbon to nitrogen ratios in red giants are a complex function of mass, metallicity, and evolutionary state \citep[see e.g.][]{MasseronGilmore2015} and their measurement may slightly depend on the inclusion of rotation (Troup et al. 2015, in prep) \footnote{We note, however, that the CNO abundances
in most of our rapidly rotating stars as measured with spectral libraries including rotation (Troup et al. 2015, in prep) are very similar (differences lower than $\sim$ 0.1 dex) to those used in this paper (Table 4). The only exceptions are KIC 11775041 and KIC 2285032, for which the CNO abundances including rotation differ by more than 0.1 dex.}. The measurement of carbon isotope ratios is more robust because these ratios are almost independent of the adopted model parameters \citep[see e.g.,][] {Garcia-Hernandez2009, Garcia-Hernandez2010}. Mixing diagnostics in combination with our other measurements may be helpful in understanding the origin of our rapidly rotating stars.}  For example, if {the stars with large rotational velocities have activated some kind of internal extra-mixing \citep[which can produce lithium, see e.g.,][] {Carlberg2012},} then we would expect very low $^{12}$C/$^{13}$C ratios in these stars; i.e., lower than the first dredge-up (FDU) values.

For each rapidly rotating star in our sample, we constructed specific ATLAS9 model atmospheres \citep{Meszaros2012} using the effective temperature (T$_{\rm eff}$), surface gravity (log $g$), metallicity ([M/H]), and CNO abundances as given in Data Release 12 {\citep[DR12;][]{Alam2015, Holtzman2015}} and with a fixed microturbulence of 2 kms$^{-1}$. Synthetic spectra using the same stellar parameters and abundances as the model atmospheres were generated with several $^{12}$C/$^{13}$C ratios using the Turbospectrum spectral synthesis code \citep{AlvarezPlez1998, Plez2012}. They were computed in air (wavelength step of 0.03 \AA) using the DR12 atomic/molecular line lists {\citep{Shetrone2015}} and fixed microturbulent velocity ($\xi$= 2 kms$^{-1}$). The synthetic spectra were then smoothed to the APOGEE resolution ($R$ = 22,500) and convolved with a rotational profile for each $v\sin i$ value (Table 2). The DR12 observed APOGEE spectra were compared to these synthetic spectra in order to estimate the $^{12}$C/$^{13}$C ratios (mostly lower limits; see Figure 8) from two spectral regions containing $^{13}$C$^{14}$N lines \citep[those around $\sim$15315 and 15355 \AA; see][]{Smith2013}. Figure 8 displays an illustrative example of the $^{12}$C/$^{13}$C fits in one star in our sample.  {We used only the} $^{13}$C$^{14}$N molecular lines because those from $^{13}$C$^{16}$O are usually contaminated with night sky emission lines. The $^{12}$C/$^{13}$C estimates for each star in our sample are given in Table 4.
\begin{figure}[h]
 \centering
\subfigure{\includegraphics[width=0.6\textwidth, clip=true, angle=270, trim= 0.3in 0.0in 0.1in 0.5in ]{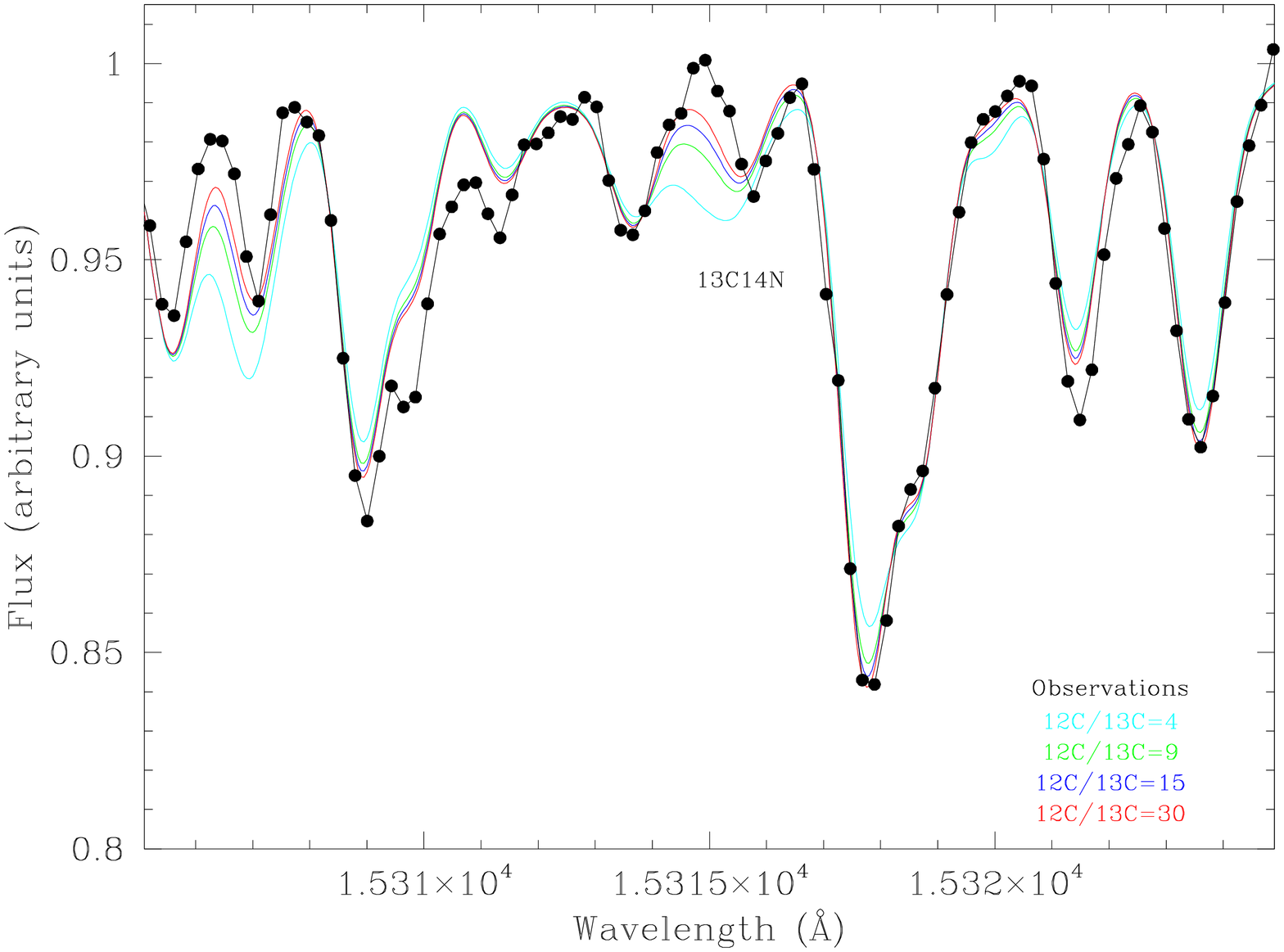}}
\subfigure{\includegraphics[width=0.6\textwidth, clip=true, angle=270, trim= 0.3in 0.0in 0.1in 0.5in ]{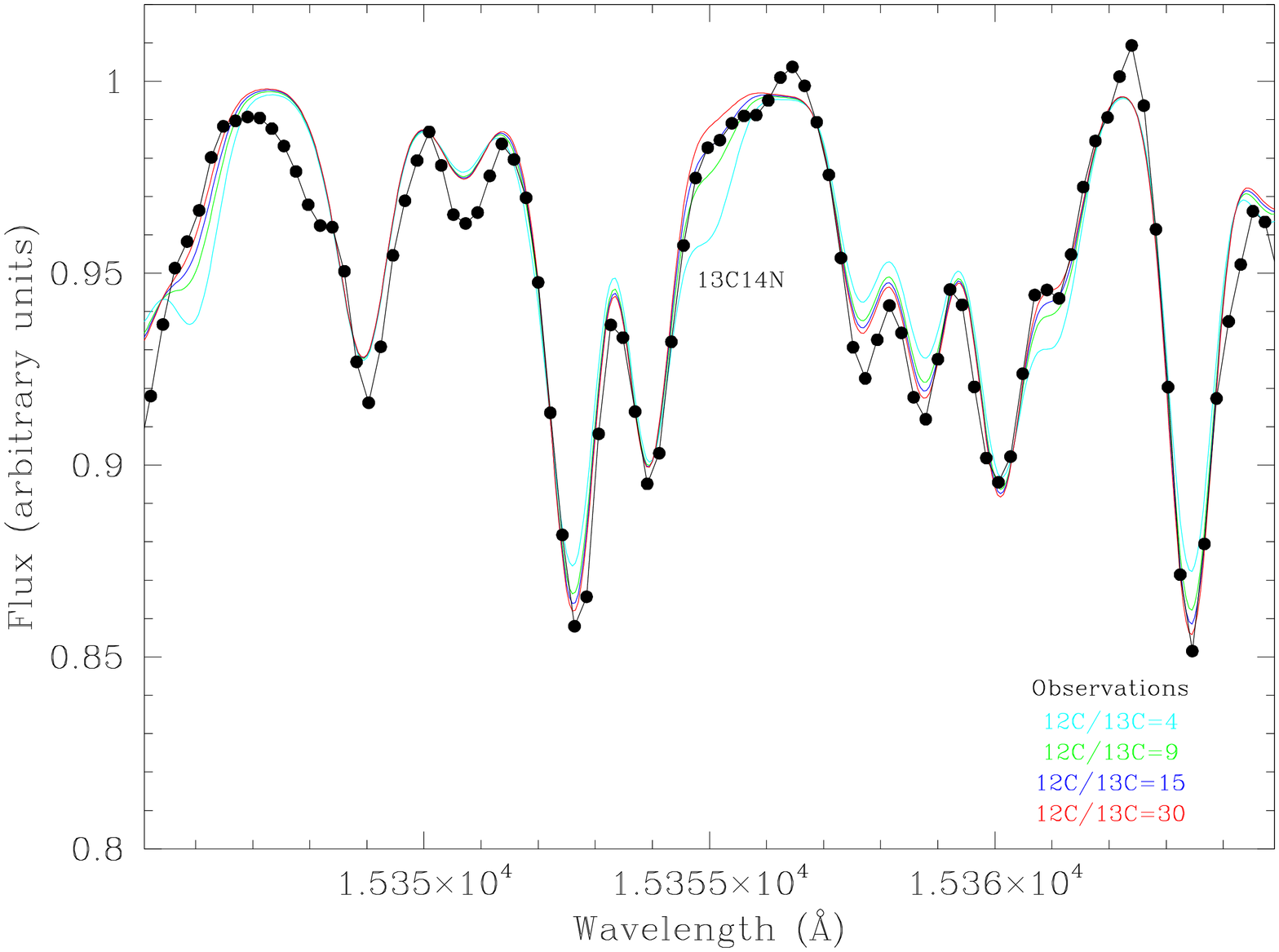}}
\caption{The comparison of the actual APOGEE spectrum {(black, $R$=22,500) of
KIC 12367827, an anomalous rotator, to synthetic spectra with various carbon
isotope ratios (colored lines) in the two wavelength regions we use to estimate
the carbon isotope ratios. In this case, the $^{12}$C/$^{13}$C ratio is estimated to be
greater than 15. Note that a few observational data points (e.g., at 15315 \AA) are
affected by imperfect telluric correction and therefore sit well above the synthetic
spectra. }}
\end{figure}
{Although the synthetic spectra are not a perfect match, the} ratios in our sample of rapidly rotating stars appear to be greater than 10-15 (see Table 4). Our measurements are lower limits because there is not much difference between synthetic spectra with $^{12}$C/$^{13}$C $>$ 10-15. Low $^{12}$C/$^{13}$C ratios ($\leq$ 10) are, however, clearly excluded. {We note that the C, N, and O uncertainties (see Table 4) are
small enough ($\sim$0.05 dex on average) that they do not significantly affect the $^{12}$C/$^{13}$C measurement. Even considering much higher CNO variations of $\pm$0.1 dex,
our conclusion about the $^{12}$C/$^{13}$C ratios being inconsistent with extra-mixing
internal to the stars is unaltered.}

This is an important clue because low $^{12}$C/$^{13}$C ratios ($<$ 10) are evidence of some extra-mixing (e.g., cool bottom processing), which is internal to the star. For example, Li-rich K giants that show very low $^{12}$C/$^{13}$C ratios (typically 5$-$7) are usually interpreted as Li regeneration internal to the star \citep[see e.g.,][]{Kumar2011}. On the other hand, higher $^{12}$C/$^{13}$C ratios (e.g., 15$-$25) are typical of stars after the first dredge-up (FDU); Li-rich K giants showing FDU $^{12}$C/$^{13}$C ratios are thus interpreted as Li production by planet accretion \citep[see e.g.,][]{Carlberg2012}.

\begin{sidewaystable}[htbp]
 \begin{adjustwidth}{-1cm}{}
\begin{tabular}{rlrrrrrrrrrl}
\hline\hline
\multicolumn{1}{l}{KIC ID} & 2Mass ID & \multicolumn{1}{l}{Log \it{g}} & \multicolumn{1}{l}{T$_{\rm eff}$} & \multicolumn{1}{l}{[Fe/H]} & \multicolumn{1}{l}{[C/H]} & \multicolumn{1}{l}{$\sigma_{[C/H]}$} & \multicolumn{1}{l}{[N/H]} & \multicolumn{1}{l}{$\sigma_{[N/H]}$} & \multicolumn{1}{l}{[O/H]} & \multicolumn{1}{l}{$\sigma_{[O/H]}$} & $^{12}$C/$^{13}$C \\ 
&& (cgs) &(K) &&&&&&&& \\ \hline
3098716 & J19044513+3817311 & 2.84 & 4802.1 & -0.379 & -0.28 & 0.06 & -0.43 & 0.04 & -0.16 & 0.04 & $>$ 15 \\ 
3937217 & J19031206+3903066 & 2.91 & 4690.2 & -0.115 & 0.02 & 0.04 & -0.13 & 0.04 & -0.06 & 0.03 & $>$ 15 \\ 
4637793 & J19035057+3946161 & 2.79 & 4587.2 & -0.080 & -0.08 & 0.04 & 0.05 & 0.04 & 0.01 & 0.03 & $>$ 15 \\ 
4937056 & J19411631+4005508 & 3.03 & 4768.9 & 0.067 & 0.07 & 0.04 & 0.26 & 0.04 & 0.10 & 0.03 & $>$ 15 \\ 
5774861 & J19043344+4103026 & 2.84 & 4616.2 & -0.076 & 0.13 & 0.04 & -0.07 & 0.04 & 0.09 & 0.03 & $>$ 15 \\ 
6501237 & J18543598+4155476 & 3.08 & 4692.2 & -0.053 & 0.00 & 0.04 & 0.05 & 0.04 & -0.01 & 0.03 & $>$ 15 \\ 
8479182 & J18564010+4430158 & 3.02 & 4695.3 & 0.007 & 0.03 & 0.04 & 0.15 & 0.04 & 0.05 & 0.03 & $>$ 15 \\ 
9390558 & J18592488+4556131 & 2.84 & 4631.4 & -0.218 & -0.04 & 0.04 & -0.18 & 0.04 & 0.01 & 0.03 & $>$ 10 \\ 
9469165 & J19341437+4605574 & 2.92 & 4756.7 & -0.589 & -0.45 & 0.07 & -0.59 & 0.04 & -0.33 & 0.04 & NC \\ 
10128629 & J19053778+4708331 & 3.02 & 4791.8 & -0.041 & -0.03 & 0.04 & 0.10 & 0.06 & -0.04 & 0.03 & $>$ 15 \\ 
10198347 & J19102813+4716385 & 3.12 & 4834.4 & -0.216 & -0.24 & 0.05 & -0.10 & 0.05 & -0.27 & 0.03 & $>$ 10 \\ 
11129153 & J19095361+4846325 & 2.91 & 4611.0 & -0.229 & -0.05 & 0.04 & -0.26 & 0.05 & 0.00 & 0.03 & $>$ 15 \\ 
11289128 & J19095233+4901406 & 2.83 & 4915.5 & -0.732 & -0.58 & 0.09 & -0.56 & 0.03 & -0.48 & 0.05 & NC \\ 
11775041 & J19491544+4959530 & 3.04 & 4914.4 & -0.489 & -0.40 & 0.07 & -0.39 & 0.02 & -0.35 & 0.04 & $>$ 10 \\ 
12367827 & J19461996+5107396 & 3.06 & 4769.0 & -0.101 & -0.03 & 0.04 & 0.09 & -9999.00 & -0.11 & 0.03 & $>$ 15 \\ 
2285032 & J19063516+3739380 & 2.72 & 4283.8 & -0.615 & -0.30 & 0.05 & -1.05 & 0.03 & -0.49 & 0.03 & $>$ 10 \\ 
2305930 & J19282563+3741232 & 2.96 & 4725.9 & -0.452 & -0.17 & 0.06 & -0.51 & -9999.00 & -0.20 & 0.04 & $>$ 10 \\ 
3955867 & J19274322+3904194 & 3.11 & 4535.2 & -0.460 & -0.46 & 0.05 & -0.46 & 0.04 & -0.43 & 0.03 & NC \\ 
4473933 & J19363898+3933105 & 3.08 & 4493.6 & -0.167 & -0.17 & 0.04 & -0.17 & 0.04 & -0.16 & 0.03 & $>$ 10 \\ 
5193386 & J19343842+4021511 & 3.46 & 4706.4 & -0.346 & -0.47 & 0.06 & -0.26 & 0.04 & -0.23 & 0.04 & NC \\ 
10293335 & J19533348+4722375 & 2.45 & 4363.4 & -0.651 & -0.58 & 0.05 & -0.62 & 0.05 & -0.62 & 0.03 & $>$ 10 \\ 
10417308 & J19460712+4730532 & 2.96 & 4734.0 & -0.472 & -0.26 & 0.06 & -0.52 & 0.04 & -0.29 & 0.04 & $>$ 10 \\ 
11497421 & J19044946+4929242 & 2.98 & 4684.5 & -0.189 & 0.08 & 0.05 & -0.23 & 0.06 & -0.04 & 0.03 & $>$ 10 \\ 
11597759 & J18554535+4938325 & 2.98 & 4658.1 & -0.190 & 0.06 & 0.04 & -0.28 & 0.03 & -0.04 & 0.03 & $>$ 10 \\ 
12003253 & J19015178+5024593 & 3.04 & 4674.4 & -0.095 & 0.05 & 0.04 & -0.06 & 0.01 & 0.00 & 0.03 & $>$ 10 \\ \hline
\end{tabular}
\end{adjustwidth}
\caption{The Data Release 12  parameters {\citep{Alam2015} used to construct the synthetic spectra} for the carbon isotope ratio analysis as well as the results of that analysis. -9999 indicates that the ASPCAP pipeline did not return an error value. Where possible, we give a lower limit on the $^{12}$C/$^{13}$C ratio for each of our anomalously and rapidly rotating stars. Stars where the synthetic spectra were not very sensitive to the carbon isotope ratio, we provide no constraint (NC). }
\end{sidewaystable}

The fact that $^{12}$C/$^{13}$C is never lower than 10 suggests that the rapid rotation sample could be the result of recent planet accretion (engulfment). We would, however, need the Li information in these stars in order to confirm this hypothesis \citep[e.g.,][] {Adamow2012}. Indeed, all of our rapidly rotating stars would be excellent targets for future high-resolution optical observations both to search for Li (i.e., by observing the Li I 6708 \AA\ line) and other planets remaining in the system (radial velocity monitoring).

\section{Interaction Rates of Low-Mass Stars}
Our calculations indicate that all stars less massive than the Kraft break (about 1.3 M$_{\sun}$) should have undetectable surface rotation rates unless they have recently interacted with a companion. Given a typical mass error of $\sim$0.1 M$_\sun$, at the 2$\sigma$ level all stars with mass estimates $<$ 1.1M$_\sun$ would be expected to be rotating slowly. We therefore assert that all stars in our sample below 1.1 M$_\sun$ rotating above our detection limit of 5 km~s$^{-1}$ are the result of a recent interaction. Of the 433 stars in this mass range, 28 of them (6.5\%) have measurable surface rotation. However, if we focus specifically on the location of the red clump (log {\it g} between 2.3 and 2.6), where stars have only recently contracted from their largest size and are thus most likely to have interacted recently with a close companion, the fraction of stars with measurable surface rotation rises to 7.6\% (26 out of 337 stars). Accounting for contamination by first ascent red giants with similar surface gravities, we suggest that at least 7\% of stars are spun up by an interaction with a companion on the upper red giant branch. We emphasize this point because recent work \citep{Mosser2012b} has indicated that the low-mass red clump stars with measurable core rotation have almost mass independent core rotation rates, which are significantly faster than expected if these stars rotate as solid bodies.  {However, given that only 24\% of giants analyzed in that work had measured core rotation rates,}  we suggest that if both the cores and surfaces of a red giant are spun up during an interaction, then the measured mean value of low-mass core rotation rates on the red clump could be biased significantly by binary interactions, and may in fact be completely disjoint from the average core rotation rates of isolated red clump stars.

\section{Conclusions}
In this work, we have quantified the expected surface rotation rates of red giant stars using stellar models and measured the projected rotation rates of the stars in the combined APOGEE-\textit{Kepler} sample.  We have identified three { rapidly rotating stars in} known eclipsing binaries and {seven} additional stars rotating unusually rapidly as well as {15} stars rotating anomalously; these rotation rates likely indicate a recent interaction. Rapid rotators represent only {0.3\%} of the sample, a number significantly lower than the 1.3 to 2.3 \% we would have expected from analyses of rapid rotation due to recent binary interactions or mergers in the field. The APOKASC seismic sample {appears to be} depleted in interaction products compared to the field. We also identify KIC 10293335 as a likely merger product, and KIC 6501237 as a possible binary system of two oscillating red giants. Finally, we note that at minimum 7\% of low-mass stars interact on the upper red giant branch, a measurement that might have significant implications for the interpretation of core rotation.

\acknowledgements{We thank Molly Gallagher, Saskia Hekker, Steve Majewski, and Benoit Mosser for helpful discussions. We also thank the referee for helpful comments and insightful questions which substatially improved the manuscript. JT, JAJ and MP acknowledge support from NSF grant AST-1211673 and NSF grant 60043121. JT and MP acknowledge support from NSF grant AST-1411685. D.A.G.H. and O.Z. acknowledge support provided by the Spanish Ministry of
Economy and Competitiveness under grant AYA-2011-27754. SM acknowledges support from the NASA grant NNX12AE17G. AS is partially supported by the MICINN grant ESP2013-41268-R and SGR 1458 (2014-2016). DS acknowledges support from the Australian Research Council. This research was supported in part by the National Science Foundation under Grant No. PHY11-25915.

Funding for SDSS-III has been provided by the Alfred P. Sloan Foundation, the Participating Institutions, the National Science Foundation, and the U.S. Department of Energy Office of Science. The SDSS-III web site is http://www.sdss3.org/.

 SDSS-III is managed by the Astrophysical Research Consortium for the Participating Institutions of the SDSS-III Collaboration including the University of Arizona, the Brazilian Participation Group, Brookhaven National Laboratory, Carnegie Mellon University, University of Florida, the French Participation Group, the German Participation Group, Harvard University, the Instituto de Astrofisica de Canarias, the Michigan State/Notre Dame/JINA Participation Group, Johns Hopkins University, Lawrence Berkeley National Laboratory, Max Planck Institute for Astrophysics, Max Planck Institute for Extraterrestrial Physics, New Mexico State University, New York University, Ohio State University, Pennsylvania State University, University of Portsmouth, Princeton University, the Spanish Participation Group, University of Tokyo, University of Utah, Vanderbilt University, University of Virginia, University of Washington, and Yale University. }
\bibliographystyle{apj} 
\bibliography{RapidRottext2}
\end{document}